\documentclass[fleqn,twoside]{article}
\usepackage{espcrc2}

\usepackage{graphicx}
\usepackage{epsfig}

\usepackage[figuresright]{rotating}

\newcommand{\AmS}{{\protect\the\textfont2
  A\kern-.1667em\lower.5ex\hbox{M}\kern-.125emS}}

\title{Very High Energy Cosmic Rays and Their Interactions}

\author{
  Ralph Engel\addressmark[FZK]\address{Forschungszentrum Karlsruhe,
    Institut f\"ur Kernphysik, 76021 Karlsruhe, Germany}
}

\begin{document}

%%%%%%%%%%%%%%%%%%%%%%%%%%%%%%%%%%%%%%%%%%%%%%%%%%%%%%%%%%%%%%%

\begin{abstract}
The investigation of high energy cosmic rays and their interactions
is a very active field of research.  This article summarizes the progress
made during the last years 
as reflected by the contributions to the XIII International Symposium on Very
High Energy Cosmic Ray Interactions held in Pylos, Greece.
\end{abstract}

\maketitle

%%%%%%%%%%%%%%%%%%%%%%%%%%%%%%%%%%%%%%%%%%%%%%%%%%%%%%%%%%%%%%%%%

\section{Introduction}

The investigation of very high energy cosmic rays and their interactions
are inherently connected subjects of astroparticle physics.
On one hand the understanding of cosmic ray interactions is needed to
study the flux, acceleration and propagation of cosmic rays.
For example, the high energy cosmic ray flux can only be measured by
linking
secondary particle cascades observed in detectors or the
Earth's atmosphere to primary particles of certain energy, mass number
and arrival direction.
Furthermore the knowledge of particle production
is needed for the interpretation of secondary particle
fluxes due to cosmic ray interactions in various astrophysical
environments.
On the other hand cosmic rays provide us with a continuous beam of high
energy particles that can be exploited for studies of interaction
physics at energies and phase space regions not accessible at
man-made accelerators.

Cosmic ray research of the last years is characterized by substantial
progress in measuring primary and secondary particle fluxes. 

Examples for new results on the primary cosmic ray flux are 
the measurements below the knee by AMS, BESS and ATIC, 
in the knee energy region by KASCADE and
TIBET, and at the highest energies by the High Resolution Fly's Eye
(HiRes) experiment. Still some of the experimental results appear
contradictory and are subject of controversial discussions.
For example, 
the results of the composition analyses of the KASCADE and EAS-TOP data
seem to be in variance with a first, preliminary analysis of
the TIBET data. Similarly, there appears to be a discrepancy between the
AGASA measurements of the cosmic ray flux above $10^{20}$\,eV and the
new HiRes data.

In addition to the measurement of the primary cosmic ray flux the most
powerful method of improving our understanding of cosmic ray physics
is the study of secondary particle fluxes.
New instruments measuring gamma-rays (CANGAROO, HESS, MAGIC, VERITAS,
and Milagro), muons and neutrinos (AMANDA, BAIKAL,
NESTOR, and ANTARES) have begun taking
data or successfully performed prototype runs. Regarding cosmic ray
physics, they are expected not
only to test models of cosmic ray acceleration and
interaction in supernova remnants and other astrophysical objects but
also to provide valuable clues on cosmic ray composition and the
characteristics of high energy
particle production.

There are many efforts to develop better models for cosmic ray
interactions or to derive information on hadronic 
multiparticle production.  The progress in this field is closely
linked to measurements of forward multiparticle production
in fixed-target and collider experiments.

One of the central problems is the
consistent implementation of the consequences of the 
steeply rising parton densities measured in deep inelastic e-p
collisions at HERA
and the indications of parton density saturation seen at the 
Relativistic Heavy Ion
Collider RHIC. RHIC data clearly demonstrate the difficulties
of extrapolating models tuned to accelerator data to higher energy or
other projectile/target combinations. Many models predicted a
secondary particle multiplicity exceeding that measured in central
Au-Au collisions by $\sim 30$\% or more. The impact of the RHIC data on
the extrapolation of cosmic ray interaction models to ultra-high energy
is still far from being understood. 

Measurements of cosmic ray showers and
secondary particle fluxes have reached a precision that they
become increasingly important in constraining hadronic interaction
models. Despite providing mainly indirect information on hadronic multiparticle
production they allow the exclusion of extreme model
extrapolations and limit exotic physics scenarios.

This article presents a summary of recent developments in the field of
very high energy cosmic ray physics and related interaction physics, 
focusing on the
contributions presented at the XIII International Symposium on Very High
Energy Cosmic Ray Interactions. The plan of the
paper is as follows. In Sec.~\ref{sec:flux} the current status of cosmic
ray flux measurements is given. Results of different
measurements are compared and their dependence on hadronic 
interaction models employed for data analysis is discussed.
The progress in modeling extensive air showers is outlined in
Sec.~\ref{sec:eas-modeling}, focussing on status and
uncertainties of high-energy interaction models. Motivated by the
current use of QGSJET as ``standard candle'' interaction model in almost
all high energy cosmic ray experiments,
uncertainties and features of interaction models are
discussed in some detail.  The importance of analyzing observables
of relevance to cosmic ray physics in experiments at current and 
future accelerators is emphasized.
Sec.~\ref{sec:exotics} gives a short update on the controversial subject
of exotic interaction features claimed to be found in emulsion
chamber measurements.
The very active field of measuring gamma-rays,
muons and neutrinos produced in cosmic ray interactions is briefly
touched upon in Sec.~\ref{sec:secondaries}. Finally, 
conclusions and an outlook is given in Sec.~\ref{sec:outlook}.

%%%%%%%%%%%%%%%%%%%%%%%%%%%%%%%%%%%%%%%%%%%%%%%%%%%%%%%%%%%%%%%%%%%%%%%

\section{Cosmic ray flux\label{sec:flux}}

For understanding very high energy cosmic rays and their sources, 
the measurement and interpretation of
the all-particle flux,
the elemental composition (including the $\gamma$-ray fraction),
the arrival direction distribution 
(large scale anisotropy, small scale clustering, and
correlation with hypothetical sources), and temporal variations
are of central importance. We will briefly discuss these topics beginning
with the highest energies. Recent reviews of the experimental situation
can be found 
in~\cite{Anchordoqui:2002hs,Haungs:2003jv,%
Watson:2003ba,Cronin:2004ye,Engel:2004ui}.

%%%%%%%%%%%%%%%%%%%%%%%%%%%
\begin{figure*}[!htb]
\centerline{
\includegraphics[width=0.8\textwidth]{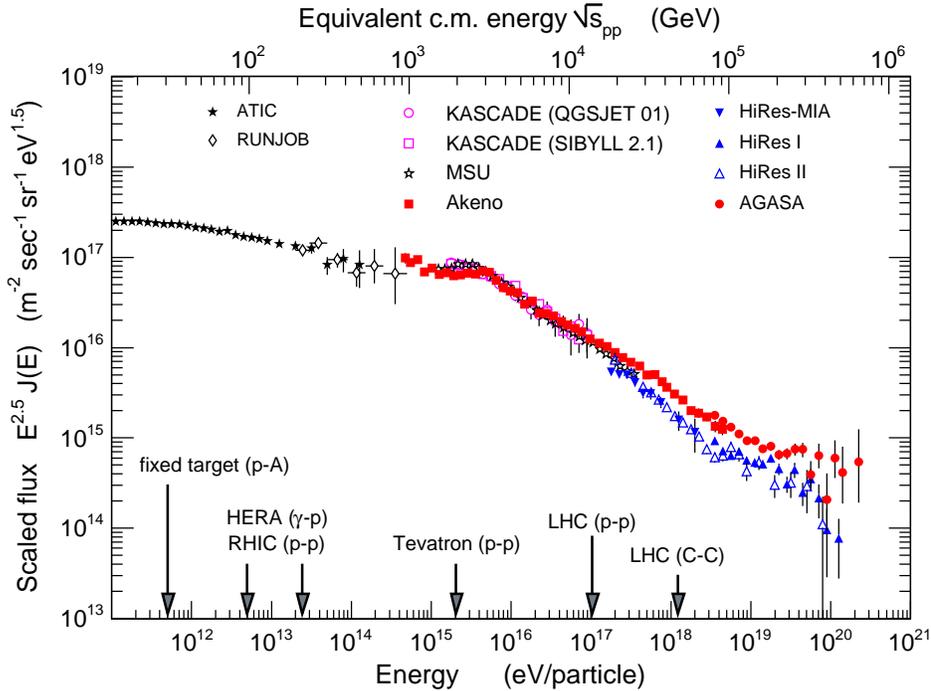}
}
\vspace*{-8mm}
\caption{
Primary cosmic ray flux scaled with $E^{2.5}$. 
Shown is a selection of recent 
measurements as discussed at this meeting together with some older data 
for reference (AGASA~\cite{Takeda:2003aa},
Akeno~\cite{Shinozaki,Nagano:1984db,Nagano:1992jz},
HiRes~\cite{Westerhoff,Abbasi:2002ta,Abu-Zayyad:2002sf},
HiRes-MIA~\cite{Abu-Zayyad:2000ay}, 
% Yakutsk~\cite{Knurenko2},
KASCADE~\cite{Haungs1,Ulrich:2004bn}, 
% Tibet~\cite{Ma1,Amenomori:2003tu},
% EAS-TOP~\cite{Navarra,Aglietta:1998te}, 
MSU~\cite{Fomin:2003tp},
RUNJOB~\cite{Shibata,Furukawa:2003dm}, ATIC~\cite{Ahn:2003cz}).
For the sake of clarity,
the all-particle fluxes from EAS-TOP~\cite{Navarra,Aglietta:1998te} and
Tibet~\cite{Ma1,Amenomori:2003tu} are not shown.
They cannot be distinguished from the others in this representation.
\label{fig:flux}
}
\end{figure*}
%%%%%%%%%%%%%%%%%%%%%%%%%%%

%%%%%%%%%%%%%%%%%%%%%%%%%%%%%%%%%%%%%%%%%%%%%%%%%%%%

\subsection{Ultra-high energy cosmic rays}

At the highest energies, the Akeno Giant Air Shower Array (AGASA) and 
the High Resolution Fly's Eye (HiRes) detectors are the 
installations with the biggest
accumulated aperture that have published flux data. 
The flux data from both experiments are shown in Fig.~\ref{fig:flux}.

First of all there is the well-known and often discussed
discrepancy between the two data sets at very high energy. Whereas the
AGASA data do not show any sign of the expected GZK
cutoff~\cite{Zatsepin} in the energy spectrum
\cite{Shinozaki,Takeda:1998ps},
the HiRes results are compatible with such a GZK signature
\cite{Westerhoff,Abbasi:2002ta}.
The statistical significance of this discrepancy above $10^{20}$\,eV
is about 2-3\,$\sigma$ \cite{DeMarco:2003ig}.
At lower energy, the overall difference between the
measurements is well within the range of the systematic errors
of both experiments \cite{Takeda:2002at,Bergman:2003aa}.
This also applies to the independent 
data set from the Yakutsk array~\cite{Knurenko2} which 
is characterized by a larger
shower-to-shower reconstruction uncertainty of 32 - 46\% as compared to
about 20\% for HiRes and AGASA. The Yakutsk array also has an
integrated aperture $\sim 40$\% smaller than AGASA.

Secondly there is a smooth transition between the HiRes and AGASA data 
and the lower energy measurements of the prototype
instrument HiRes-MIA \cite{Abu-Zayyad:2000ay} 
and the Akeno air shower array \cite{Nagano:1984db,Nagano:1992jz},
respectively. This
might indicate a systematic bias in one or both measurement techniques
that could be related to the simulation of ultra-high energy air
showers.

To estimate the elemental composition of ultra-high energy cosmic rays 
(UHECR), both AGASA and 
HiRes have analyzed their data in terms of a two-component proton/iron
composition hypothesis. 

The HiRes analysis is based on the measurement
of the depth of shower maximum. Using the hadronic
interaction model QGSJET \cite{Kalmykov92e,Kalmykov97a} for
interpreting the data they find a transition to a light 
composition \cite{Abu-Zayyad:2000ay}
that remains proton dominated (80\% protons) between $10^{17}$\,eV and 
$10^{19.3}$\,eV \cite{Westerhoff,Abbasi:2004nz}. 
Preliminary results of the re-analysis of the AGASA muon density data
measured with the Akeno muon detectors show also a transition to a
proton dominated composition. In contrast to the HiRes results,
the transition is found to 
occur gradually over a large energy range. From about 50\% iron fraction
at $10^{17.5}$\,eV the iron contribution drops below 20\% at $10^{19}$\,eV
\cite{Shinozaki,Shinozaki:2004nh}. Similar to the HiRes analysis,
the Yakutsk composition measurement~\cite{Knurenko1} is related to the primary
mass sensitivity of the shower depth of
maximum. The Yakutsk group find a light composition of
about 70-80\% proton and helium in the energy range $5\times 10^{17} -
5\times 10^{18}$\,eV~\cite{Knurenko1}.

The most natural interpretation of the changing composition
would be the transition from Galactic to extragalactic
cosmic rays. The transition energy would be somewhere between $10^{17}$
and $10^{19}$\,eV. Whereas HiRes data are consistent with the
interpretation that the ``ankle''
in the cosmic ray spectrum is already a signature of the GZK cutoff
($e^+e^-$-pair production of protons with photons of the CMB)
\cite{Bergman:2003wx,Berezinsky:2005cq}, AGASA
data favour an interpretation of the ankle as transition region between
Galactic and extragalactic cosmic rays.

It is clear, however, that these composition measurements are
strongly model dependent as there is a large theoretical uncertainty in
predicting electron and muon shower sizes as well as the
depth of shower maximum for hadron-induced
showers \cite{Drescher2,Ostapchenko2,Pierog}. 
An interpretation based on SIBYLL \cite{Fletcher94,Engel99a} or
neXus \cite{Drescher:2000ha} 
gives a heavier composition: about 30 and 50\% iron, respectively (see
also Fig.~\ref{fig:xmax-models}). 
It is unclear which of the model predictions is more realistic and also
whether the range of predictions exhausts the range of the
theoretical uncertainties (see discussion in Sec.~\ref{sec:eas-modeling}).
Moreover there are signs of inconsistencies \cite{Watson,Watson:2003ba}. 
For example, the muon densities
measured for the same showers as used in the depth of maximum analysis
of the HiRes-MIA data
are similar to or even exceed those expected for iron primaries
\cite{Abu-Zayyad:2000ay}.
Furthermore investigations based on mass-sensitive 
observables such as shower disk thickness and shape of the
lateral distribution also indicate a 
heavier composition ($\sim 80-90$\% iron) \cite{Ave:2003ab,Dova:2004nq}. 

Limits on the primary photon fraction were given by AGASA based on 
the investigation of showers with muon density information. The fraction
of photon-induced showers is found to be smaller than 28\% (67\%) at
energies greater than $10^{19}$ ($10^{19.5}$) eV at 95\% CL.
A recently developed method
of comparing shower-by-shower measurements
with theoretical expectations for photon-induced showers \cite{Homola} 
allows one to
derive a limit at even higher energy where the statistics is very
sparse: less than 65\% of all showers with energies above 
$1.25\times 10^{20}$\,eV are initiated by photons (95\% CL)
\cite{Risse:2005jr}.

Given the limited statistics accumulated until now, the arrival
direction distribution of UHECR appears isotropic. There are a number of
cosmic rays forming arrival direction multiplets in the AGASA data
set (57 events with $E>4\times 10^{19}$\,eV: 6 doublets, 1 triplet) 
\cite{Shinozaki,Teshima:2003ab}. 
The statistical
significance of this small scale clustering is subject to controversial
discussion. If the clustering were found with ``a priori'' chosen values
for energy
threshold and separation angle, i.e.\ without performing a scan in
energy threshold and separation angle, the chance probability would be
less than $10^{-4}$. Assuming that such a scan was
performed, the chance probability would increase to
about $0.3$\% \cite{Finley:2003ur}. 

The exposure of HiRes in stereo mode has not yet reached that of
AGASA\footnote{Viewing showers in monoscopic mode HiRes I has reached a
higher accumulated aperture than AGASA for energies above $3\times
10^{19}$\,eV.  However, due to the limited and highly asymmetric angular
resolution, the HiRes I mono data set is not suited for studying small
scale clustering.}. There are 27 events detected in stereoscopic mode
above $4\times 10^{19}$\,eV. Using this data set in an autocorrelation
analysis no significant small-scale clustering is found.  Adding the
HiRes stereo data set to that from AGASA only one additional pair is
found. The clustering found in the combined data set is estimated to have a
chance probability of 1\%.

The search for correlations with astrophysical sources is hampered by the
incompleteness of catalogs and related object detection and 
selection biases. Assuming that the
source distribution follows that of the luminous matter in the universe
it is natural to expect a correlation with the supergalactic plane
\cite{Stanev:1995my}. No such correlation is found in the
current data sets. There are a number of correlations claimed between the
AGASA and Yakutsk data sets and BL Lacs
\cite{Tinyakov:2001nr,Gorbunov:2002hk,Tinyakov:2001ir}. 
With an angular resolution of
about $0.7^\circ$, the HiRes stereo data set is ideally 
suited for such studies.
Using the same astrophysical objects no correlation with the HiRes
events is found for energies above $2.4$ and $4\times 10^{19}$\,eV.

Recently new indications
of a correlation with BL Lacs were found if the energy threshold for
comparison is lowered to
$10^{19}$\,eV (10 out of 271 showers have arrival directions within
$0.8^\circ$ of one of the 156 selected BL Lacs from the Veron catalog)
\cite{Gorbunov:2004bs}. 
The chance probability of finding such a correlation in an un-correlated
data set is estimated as
0.1\%, but 
only a new, independent data set will allow to assess the significance
unambiguously. Taken at face value, the correlation would have to be
interpreted as a small fraction of neutral particles in UHECR.

Currently there are two large-aperture detectors for UHECR in
construction, the Pierre Auger Observatory
\cite{Kampert,Auger}
and the Telescope Array (TA)~\cite{Fukushima:2003ig,TA}. 
Both detector concepts employ the hybrid detection technique of
measuring air showers with surface detectors (Auger: water Cherenkov
tanks, TA: plastic scintillators) and fluorescence telescopes.
The hybrid technique will allow a good energy calibration of showers
measured with surface detectors and improve the ability of composition
measurements.

After testing the detector design with an 
engineering array~\cite{Abraham:2004dt} the construction of
the southern Auger observatory in Malargue, Argentina
is in full progress \cite{Kampert}. 
At the time of writing this article about 800
of the planned 1600 surface detector stations
and 50\% of the fluorescence telescopes are completed. Already during the
construction phase the integrated aperture of Auger has reached that
of 10 years of data taking with AGASA. It is planned to build a similar
observatory in the northern hemisphere to obtain nearly uniform full sky
coverage.

%%%%%%%%%%%%%%%%%%%%%%%%%%%%%%%%%%%%%%%%%%%%%%%%%%%

\subsection{Knee energy region}

In contrast to previous measurements in the knee energy region
(for example, see the compilations
in \cite{Hoerandel:2004gv,Haungs:2003jv,Swordy:2002df})
the recent all-particle flux measurements by KASCADE \cite{Haungs1}, 
EAS-TOP \cite{Navarra},
and Tibet \cite{Amenomori:2003tu} agree within 10\% with each other. 
The knee is at about $3\times 10^{15}$\,eV and no deviation
from a broken power law with a smooth transition region is found within
the current experimental resolution.

Concerning the elemental composition, the situation is less clear. There
is increasing evidence for a transition from a mixed to a heavy
composition with increasing energy. However, the detailed change of
composition through the knee energy range is still unclear.

All experimental results discussed
at the meeting show a trend towards a heavy composition with increasing
energy (KASCADE~\cite{Haungs1}, Tibet~\cite{Ma1}, EAS-TOP~\cite{Navarra},
and SPASE/AMANDA~\cite{Karle}).
However, it is difficult to find observables
that demonstrate this transition beyond doubt as all composition studies
depend very much on the hadronic interaction models used for data
interpretation and alternative explanations might be possible, although
unlikely (see, for example, \cite{Petrukhin}). 
Probably the least model dependent analysis is that of
muon-rich and muon-poor showers by the KASCADE
Collab.~\cite{Antoni:2002au},
which demonstrates that the knee is mainly caused by a
disappearance of the light (i.e.\ muon-poor) flux components.

%%%%%%%%%%%%%%%%%%%%%%%%%%%%%%%%%%%%%%%%%%%%%
\begin{figure}[htb!]
\centerline{
\includegraphics[width=7cm]{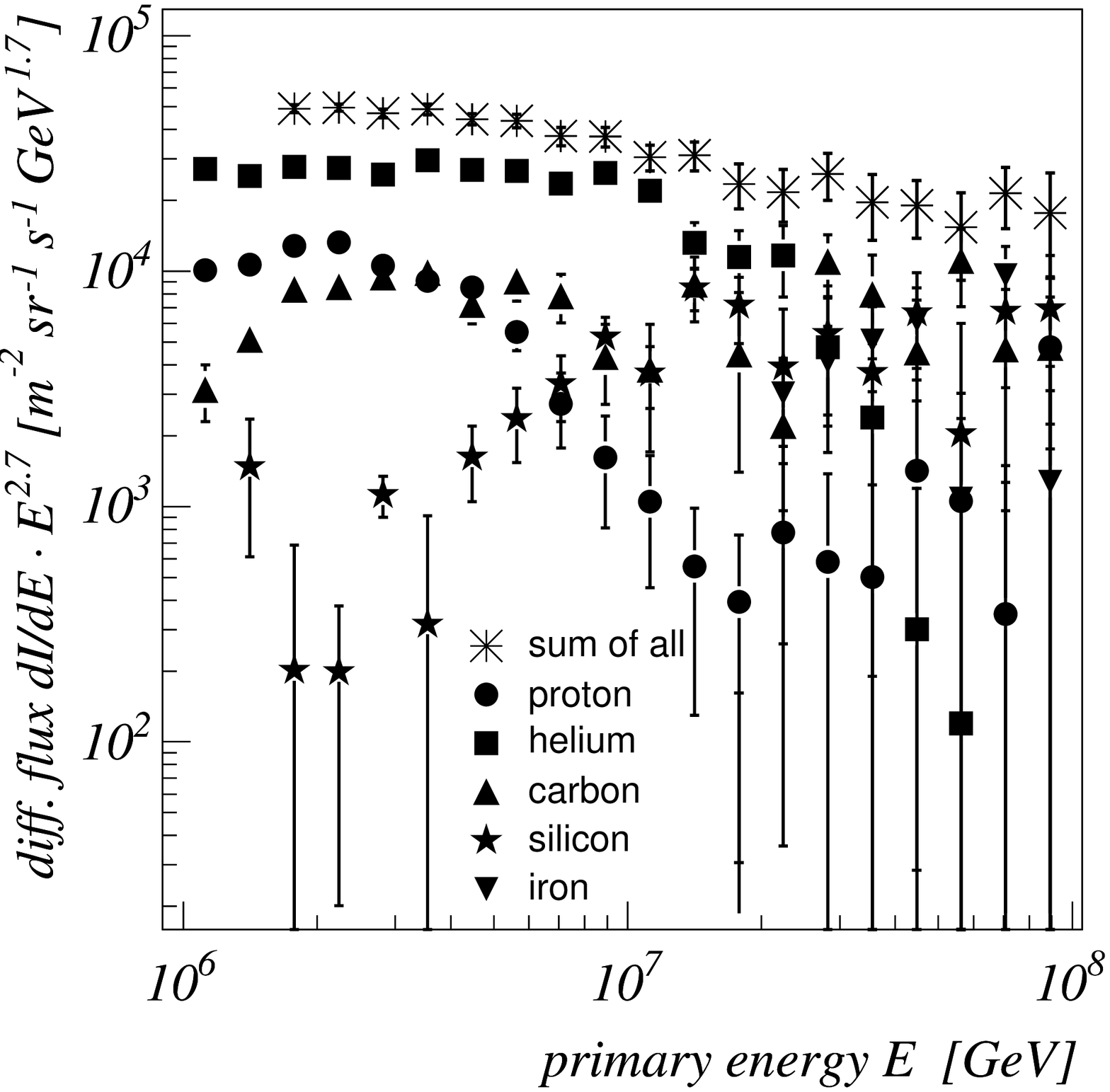}
}
\centerline{
\includegraphics[width=7cm]{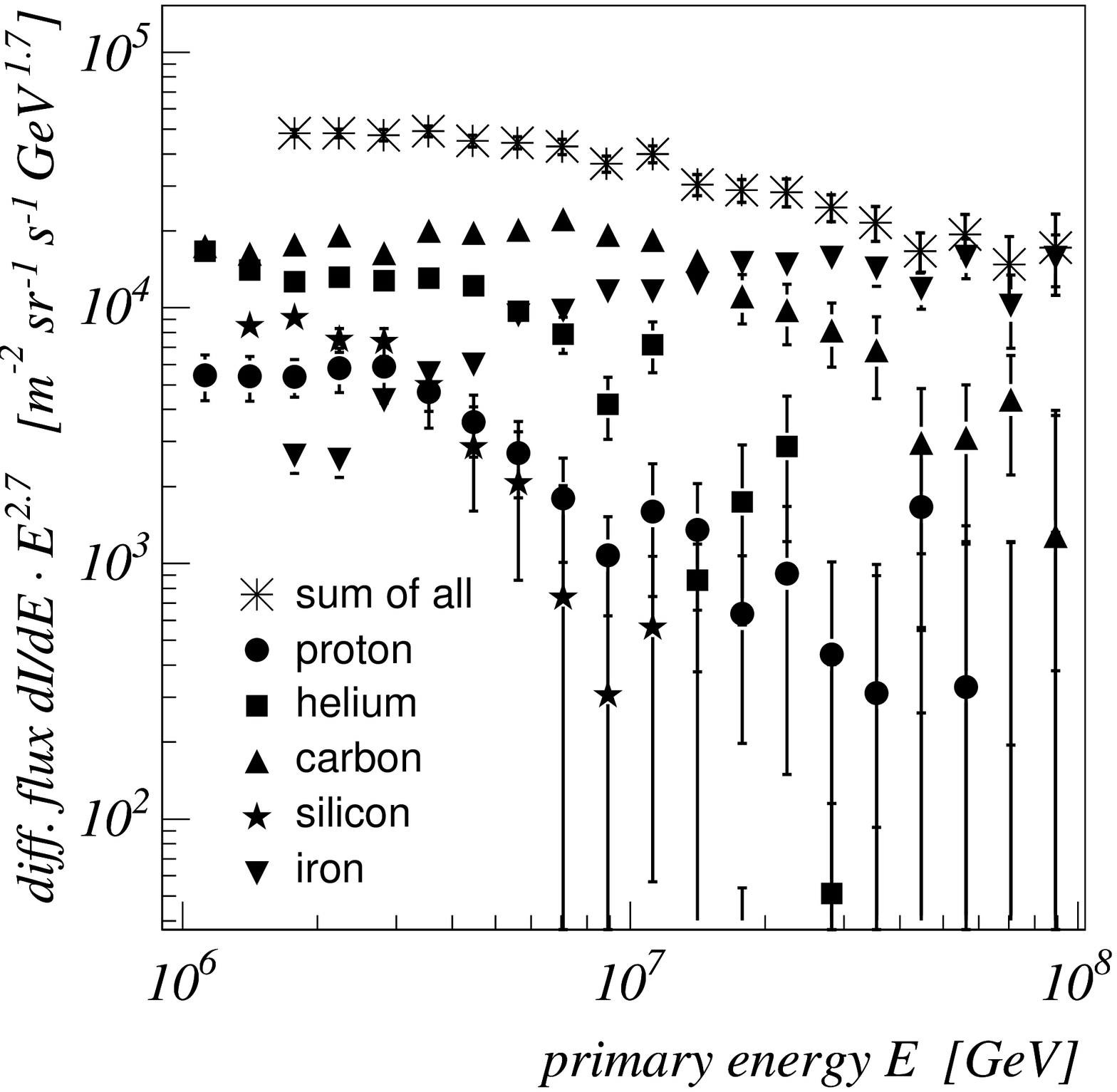}
}
\vspace*{-8mm}
\caption{
Flux derived for 5 elemental groups from KASCADE data
\cite{Haungs1,Ulrich:2004bn}.
The top panel shows the results obtained with QGSJET 01, 
the bottom panel those with SIBYLL 2.1.
\label{fig:KASCADE-comp}
}
\end{figure}
%%%%%%%%%%%%%%%%%%%%%%%%%%%%%%%%%%%%%%%%%%%%%

The high-statistics data of multi-detector 
setup KASCADE \cite{Antoni:2003gd}
allow the analysis of the correlated 
muon ($E_\mu > 240$\,MeV) and electron sizes of showers in terms of
5 mass groups \cite{Haungs1,Ulrich:2004bn}. Fig.~\ref{fig:KASCADE-comp}
shows the results for two hadronic interaction models, QGSJET and
SIBYLL. In this analysis, the derived all-particle flux turns out to be
almost independent of the used hadronic interaction model. However, the
different elemental fluxes vary strongly. For example, using QGSJET the flux
appears dominated by helium below the knee with no significant iron
contribution. The SIBYLL-based interpretation favours instead helium and
carbon below the knee and a small but significant fraction of iron
primaries is needed.
Both models do not provide a fully consistent description of the KASCADE
$N_e$-$N_\mu$ data. The found deviations underline the high statistical
accuracy of the KASCADE data and show the need of improving hadronic
interaction models.

The interpretation of the KASCADE data with both models shows that
the flux of the light components exhibits a break in the power law at
different energies with  lighter elements having a lower break energy.
No spectral break is found for iron in the considered energy range. 
One of the central questions is that of the scaling of the break
energies. Acceleration and propagation models for the knee typically
predict rigidity-dependent scaling whereas models with new particle
physics lead to mass-dependent scaling.
Unfortunately, the strong had.\ interaction
model dependence does not allow us to 
draw conclusions on
a possible mass- or rigidity-dependent scaling of the break energies.

The composition measurement by the EAS-TOP
Collab.~\cite{Navarra,Aglietta:1998te} 
is based on 3 elemental groups and uses electron and GeV-muon
data.
It shows the same qualitative behaviour
of the elemental groups as seen in the KASCADE analysis. The iron-like
mass group does not exhibit a break in the power spectrum and seems to
have a harder power law index than the light component. The knee appears
to be caused by elements in the mass range of proton and helium with a
break energy of about $3.5\times 10^{15}$\,eV.

Some of the air showers measured with the EAS-TOP and SPASE arrays 
produce high-energy muon bundles that can be detected with MACRO 
($E_\mu > 1.3$\,TeV) and AMANDA ($E_\mu > 300$\,GeV),
respectively. Again, analyses of the coincidence data sets show a
transition from a mixed to a heavy composition
\cite{Navarra,Aglietta:2003hq,Karle,Ahrens:2004nn}.

The preliminary and statistically limited 
measurement of the proton flux by the Tibet AS$\gamma$
Collab.~\cite{Amenomori:2003tk} 
seems to be at variance with KASCADE and EAS-TOP
results. Over the energy range from $2\times 10^{14}$ to $10^{16}$\,eV
the proton spectrum is found to follow a power law with the index $3.14
\pm 0.10$. By comparing the Tibet proton spectrum with that of RUNJOB and
JACEE a much lower break energy of $\sim 5\times 10^{14}$\,eV is inferred. 

The reasons for the discrepancy between Tibet and KASCADE/EAS-TOP data
is not yet understood. However, it should be noted that both KASCADE and
EAS-TOP employ in their analysis the electron-muon size correlations in
showers whereas the Tibet measurement is based on a neural net
analysis of a number of shower observables that combine emulsion chamber
information with scintillator data: 
$N_\gamma$ (multiplicity of a family), 
$\sum E_\gamma$ (energy sum of a family), $\langle
R_\gamma \rangle$ (mean lateral spread of
a family), $\langle E_\gamma \cdot R_\gamma \rangle$
(mean lateral spread of the family energy), and
$N_e$ (shower size) \cite{Amenomori:2003tk}. Furthermore 
all three experiments are at different altitudes and probe different
stages of shower evolution. Tibet is 
at a vertical depth of 606 g/cm$^2$, EAS-TOP at 820
g/cm$^2$ and KASCADE at 1020 g/cm$^2$. The use of different air shower
simulations can be ruled out as all experiments now apply CORSIKA
\cite{Heck98a}.

Due to the large statistical errors, the direct flux measurements of
individual mass groups by JACEE \cite{Asakimori:1998aa}, RUNJOB
\cite{Shibata,Furukawa:2003dm} and ATIC \cite{Ahn:2003de} do not 
yet impose strong
constraints on the air shower data. All composition data discussed here
are compatible with possible extrapolations of direct measurements at lower
energy.

In cosmic ray models that explain the knee by propagation effects (leakage from
our Galaxy), an increasing dipole anisotropy of the shower arrival directions 
is expected (i.e.~\cite{Candia:2003dk}).
Analyzing $2\times 10^7$ showers in the knee energy range
the KASCADE
group do not find any significant anisotropy signal \cite{Antoni:2003jm}. 
Similarly, no cosmic ray point sources are seen \cite{Antoni:2004sc}. 

Being located at higher altitude, the Tibet array has a much lower
energy threshold. Again, no Galactic anisotropy was found in the Tibet
AS$\gamma$ data. However, using more than $5\times
10^{9}$ showers with $E>3\times 10^{12}$\,eV, the Tibet AS$\gamma$
Collab.\ could detect the 
dipole anisotropy due to the orbital motion of the Earth
around the Sun (Compton-Getting effect) with an amplitude of about 0.1\% 
\cite{Ma1,Amenomori:2004bf} (see also \cite{Aglietta:1996sz}).

%%%%%%%%%%%%%%%%%%%%%%%%%%%%%%%%%%%%%%%%%%%%%%%%%%%%%%%%%%%%%%%%%%%%%%%

\section{Modeling of cosmic ray interactions and EAS\label{sec:eas-modeling}}

Given the dependence of the cosmic ray flux and composition measurements
on the understanding and modeling of hadronic interactions of cosmic
rays and their secondary particles, it is natural to assume that
discrepant results discussed in the previous section 
can be, at least to some extent, traced back to the use
of different models for data interpretation.

Several groups have shown that an analysis of the same
data set with different hadronic interaction models can lead to a wide
range of different results (see, for example,
\cite{Aglietta:1998te,Ulrich:2004bn}). 
This means that the use of the same shower
simulation model is pre-requisite for a fair comparison of the results
of different experiments. 

Another important aspect of inter-experiment comparison is the use of
sufficiently realistic and accurate shower simulations. For example, 
using the same model for
shower evolution, one can obtain different interpretations of the data
if different observables of the showers are considered \cite{Antoni:2001pw}.  
Therefore experiments might arrive at contradicting conclusions even if
the same shower simulation tools are used.

The largest uncertainty in EAS simulation stems from the unknown
characteristics of hadronic multiparticle production
\cite{Knapp96a,Knapp:2002vs}. As has been realized during the last
years, also interactions at intermediate energies can contribute
significantly to this uncertainty, though to a smaller extent
\cite{Engel99c,Drescher:2002vp,Heck:2003br}. In addition there are
uncertainties coming from the treatment of electromagnetic interactions
and differences in details of particle transport and decay
implementations \cite{Knapp:2002vs}.

Motivated by different models of hadronic multiparticle
production, three energy regions are distinguished. At very low energy
(close to the particle production threshold up to a few GeV) particle
production is characterized by the production and decay of resonances.
Knowing all resonances and their decay branching ratios allows one to
construct a rather complete model for the interaction cross section and
hadronic final states. For example, the codes HADRIN~\cite{Haenssgen86a} and 
SOPHIA~\cite{Muecke00a} have many resonance channels tabulated. 
The intermediate
energy range up to about $10^3$ GeV can be well understood in a model
that describes particle production on the basis of the fragmentation of
two color strings (i.e. older versions of FLUKA~\cite{Fasso01a}). At
energies above $10^3$ GeV minijet production and multiple parton-parton
interactions become important and require again a different modeling.
The most frequently used high energy models are
DPMJET~\cite{Ranft95a}, neXus~\cite{Drescher:2000ha}, 
QGSJET~\cite{Kalmykov97a}, and
SIBYLL~\cite{Fletcher94}.

In the following we will summarize some important recent developments
in modeling hadronic interactions and
related activities of measuring hadron production in accelerator
experiments,
and will discuss new trends in air shower simulation.

%%%%%%%%%%%%%%%%%%%%%%%%%%%%%%%%%%%%%%%%%%%%%%%%%%%%%

\subsection{Hadronic interactions at high energy}

%%%%%%%%%%%%%%%%%%%%%%%%%%%%%%%%%%%%%%%%%%%%%%%%
\begin{figure*}[htb!]
\centerline{
\includegraphics[width=0.7\textwidth]{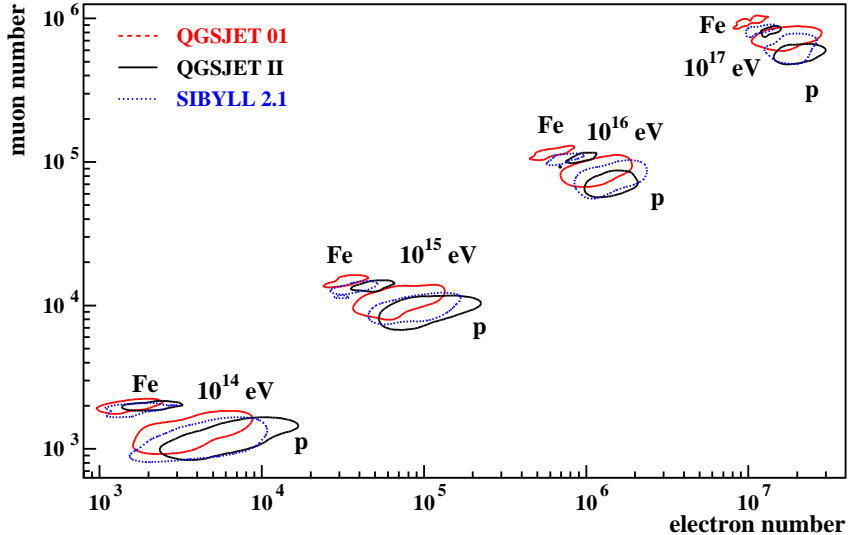}
}
\vspace*{-8mm}
\caption{
Electron-muon size correlation for showers simulated with CORSIKA
\protect\cite{Heck-Ostapchenko-private}.
Predictions of the old and new versions of QGSJET \cite{Ostapchenko2}
are compared with SIBYLL 2.1 \cite{Engel99a}.
\label{fig:ne-nmu-plot}
}
\end{figure*}
%%%%%%%%%%%%%%%%%%%%%%%%%%%%%%%%%%%%%%%%%%%%%%%%

There are basically four central assumptions that characterize a model's
high energy extrapolation
\cite{Engel03a,Alvarez-Muniz:2002ne,Ostapchenko:2003sj}:
\begin{itemize}
\item[(i)]
size and energy dependence of the QCD minijet cross section,
\item[(ii)]
distribution of partons in transverse space (profile function),
\item[(iii)]
scaling of leading particle distributions or scaling violation, and
\item[(iv)]
treatment of nuclear effects (semi-superposition model, Gribov-Glauber
approximation, increased parton shadowing, etc.)
\end{itemize}
It is beyond the scope of this article to discuss all models and their
differences.
We shall concentrate here on aspects relevant to p-air interactions
and consider only the two most frequently applied models, QGSJET and
SIBYLL. Apart from the treatment of nucleus-nucleus collisions, QGSJET
and SIBYLL differ mainly in the first two 
points \cite{Stanev,Alvarez-Muniz:2002ne}. 

Since version 2.1, ``post-HERA'' 
parton densities are
used in SIBYLL 
for calculating the minijet cross section whereas QGSJET 01 (and the
earlier version QGSJET 98) was developed with older, ``pre-HERA'' parton
densities. Another important difference between the models is the
treatment of the minijet transverse momentum cutoff needed to restrict the
minijet calculation to the perturbative QCD domain. In QGSJET 01 an
energy-independent, constant value of 2\,GeV is used. 
The SIBYLL authors
implemented an energy-dependent transverse momentum cutoff whose value is
similar to that of QGSJET 01 at low energy but increases to about 
8\,GeV at $10^{20}$\,eV \cite{Engel99a}. This was needed as
``post-HERA''
parton density functions predict at ultra-high energy
gluon densities that lead to overlap of individual gluon wave 
functions in a proton for
a transverse momentum cutoff as low as 1.5 GeV (see \cite{Gribov83}, a
recent review on this subject is \cite{Mueller:2005me}). 
In this phase space region
non-linear evolution equations have to be used to describe parton
densities. The expected size of the non-linear corrections is
theoretically not understood and subject of intense research
\cite{Erice04}. There are models that predict an early and 
total saturation of the gluon
density (color glass condensate \cite{Iancu:2003xm}) and others with
moderate changes. 

Many experiments are searching for signs of deviations from linear
parton density evolution equations or gluon density saturation. 
Although HERA and RHIC are colliders with CMS
energies of $\sqrt{s}\sim 200$\,GeV, corresponding 
to only $2\times 10^{13}$\,eV, they are currently the best instruments
for
studying saturation effects. At HERA, parton densities of quarks 
are measured directly and, through scaling violations, that of gluons
derived. HERA data can be described assuming parton density
saturation, for example, in terms of the Golec-Biernat--W\"usthoff model
\cite{Golec-Biernat:1998js}, or applying perturbative QCD without
any non-linear effects \cite{Forshaw:2004rn,Abramowicz99a}.
At RHIC, parton densities cannot be measured directly. However, the
scaling of jet rates and other observables
with the number of participating nucleons
depends on the assumptions on the number of partons in the very
gluon-dense environment of a heavy nucleus.
Many aspects of RHIC data indicate strong deviations from
naive parton model predictions \cite{Lange,Levin:2004ak}
but the energy range of the collisions is too limited to show
unambiguously that the effects observed so far 
are requiring parton density saturation \cite{Steinberg:2004ii}.

The problem of using ``post-HERA'' parton densities for extrapolating 
hadronic interactions to ultra-high energy $\sqrt{s} \sim 500$\,TeV
within the Quark-Gluon Strings Model \cite{Kaidalov:1983vn}
is addressed in the new version of QGSJET, called QGSJET-II
\cite{Ostapchenko1}. Different
from the SIBYLL approach, the transverse momentum cutoff is kept
energy-independent. This is achieved by introducing non-linear effects
for partons below the perturbative scale, equivalent to non-linear
evolution equations. As one cannot speak of individual partons in the soft,
non-perturbative domain, these non-linear effects are implemented as
multi-pomeron interactions (enhanced pomeron graphs
\cite{Kaidalov86b-e}), which are summed
to all orders. Other important improvements are the treatment of
diffraction dissociation. Whereas the old version of QGSJET had a fixed
ratio of diffractive to elastic cross sections (grey disk limit, see
also \cite{Luna:2004eg}), the
new version approaches at high energy the black disk limit (see
discussion in \cite{Engel03a}).
Furthermore QGSJET-II was tuned to better describe the
secondary particle multiplicity at low collision energy. Although the
latter changes are more of technical nature they still are
important. The old version
predicted at low energy too high a pion multiplicity, a possible reason for
higher GeV-muon multiplicities obtained for EAS than found with other models
\cite{Engel:2002id,Heck:2003br}. It is clear that QGSJET-II is a much more
theoretically consistent model than the previous versions.

There are a number of important consequences from these changes in QGSJET.
First of all the low-$x$ extrapolation of the parton densities becomes
less steep than naively expected in linear perturbative QCD.
Secondly 
there are changes of the effective parton densities acting
in hadron-hadron collisions in dependence of the projectile and target
mass number $A$ (suppression for large $A$). 
This leads to the violation of the superposition approximation
even for fully inclusive 
observables.\footnote{In the simplest version of the superposition
model, an
iron-induced shower is equivalent to 56 proton induced showers 
having 1/56 of the
shower energy. The superposition model is expected to be valid for
inclusive observables.}
Thirdly the fluctuations in inelasticity are considerably reduced. The
leading particle distributions are now qualitatively similar to those of 
SIBYLL and DPMJET \cite{Heck-private}.

The impact of the QGSJET improvements on air shower predictions is 
currently under investigation \cite{Heck-private}. For example,
Fig.~\ref{fig:ne-nmu-plot} shows the electron-muon size correlation for
vertical EAS at sea level \cite{Heck-Ostapchenko-private}. 
The contours are iso-lines of the correlation function at half maximum.
For a given energy the number of muons is reduced significantly. At the
same time 30-40\% more electrons are expected at detector level in the
knee energy region. 
The interpretation of data from experiments
that utilize electron and muon numbers as composition sensitive
observable (e.g.\ KASCADE, EAS-TOP) will change towards a 
heavier composition if QGSJET-II is used. 

At the highest energies, the predictions of the electron numbers of the 
old and new QGSJET versions are not very different.
Still it can be expected that the energy calibration of experiments like
AGASA would have to be revised downward though detailed simulations are
needed to estimate by what amount. Experiments like Auger~\cite{Kampert} 
will be much more
sensitive to the modifications: the predicted muon density at 600 to
1000\,m from the shower core is reduced by $\sim 30$\%.
Using QGSJET-II also the interpretation of fluorescence
measurements will change and shift towards a heavier composition as
the mean depth of maximum is increased by $10$\,g/cm$^2$ 
and $20$\,g/cm$^2$ at $10^{20}$\,eV for proton and iron showers,
respectively \cite{Ostapchenko2}.

Different assumptions on the QCD minijet cross section, 
i.e. calculated with or without saturation, 
lead to enormous differences in the model extrapolations \cite{Engel03a}.
The most striking example is the 
secondary particle multiplicity in p-air collisions.
At ultra-high energy, QGSJET 01
predicts a more than 3 times higher secondary particle multiplicity
than SIBYLL \cite{Knapp96a,Alvarez-Muniz:2002ne}. 
Due to the implementation of new parton densities,
the multiplicity of QGSJET-II is even higher than that of QGSJET 01 at
ultra-high energy.  
However, these differences are of
very much reduced importance for the evolution of air showers as most of
the secondary particles have a very small energy. 
Of greater importance are two model characteristics 
that are indirectly linked to the minijet cross section:
the total p-air and $\pi$-air cross sections 
and the distribution of leading secondary particles.

%%%%%%%%%%%%%%%%%%%%%%%%%%%%%%%%%%%%%%
\begin{figure}[htb!]
\centerline{
\includegraphics[width=7.5cm]{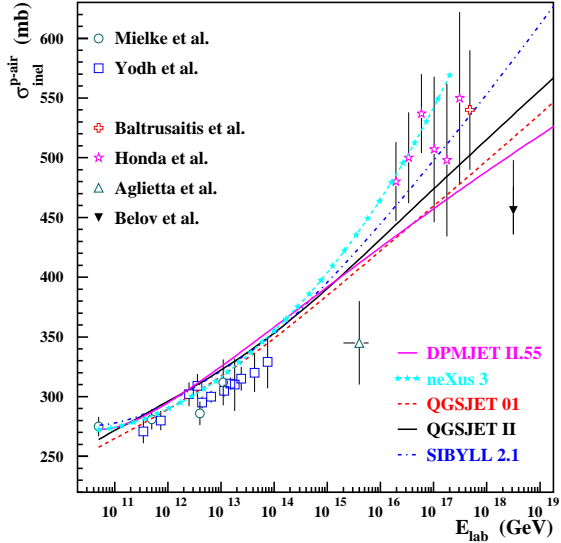}
}
\vspace*{-8mm}
\caption{
Compilation of p-air production cross sections and model predictions (from
\cite{Knapp:2002vs}, modified and updated \cite{Heck-private}). 
\label{fig:p-air-cs}
}
\end{figure}
%%%%%%%%%%%%%%%%%%%%%%%%%%%%%%%%%%%%%%

The total cross section of a model depends at high energy mainly on the
minijet cross section and the transverse profile function which is a
measure of how partons are distributed in a hadron \cite{Alvarez-Muniz:2002ne}. 
It is the difference
of the assumed profile functions and minijet cross sections
and the lack of data to distinguish
between these assumptions that lead to widely varying
high-energy cross section extrapolations (see Fig.~\ref{fig:p-air-cs}).
Cross section measurements over a wide range of energy would help to
reduce the model ambiguities. Although hadron-air cross sections are
needed for simulation, p-p and p-$\bar{\rm p}$ cross section
measurements are also of great interest. Using the Gribov-Glauber 
approximation, cross sections with air can be estimated using 
nucleon-nucleon cross section data.
At the moment, due to the
contradicting total p-$\bar{\rm p}$ cross section measurements at Tevatron
\cite{Rapidis,Albrow99a}, 
possible theoretical extrapolations are experimentally 
not very much restricted \cite{Engel03a,Zha:2003bt}. Therefore the
planned total cross section measurement with the TOTEM/CMS
detector combination at the LHC \cite{Orava,Avati:2003qj}, corresponding
to an equivalent energy of about $10^{17}$\,eV,
will be of outstanding importance.

All current high energy interaction models are based on the implicit
assumption that leading particle distributions scale
with energy. The leading particle distributions
are tuned to low-energy data and 
change at high energy only due to energy-momentum conservation
effects as the energy has to be shared between the leading particles and
the increasing bulk of low-energy secondary particles.
This assumption seems to describe the very sparse data we have on
leading particle production up to HERA energy ($\sim 2\times
10^{13}$\,eV) \cite{Engel:1998hf}. However, there are theoretical
arguments that the leading particle distributions will change
drastically at ultra-high energy \cite{Frankfurt:1997ij}.

If parton density saturation indeed occurs at cosmic ray energies, 
a collision can be viewed as black disk scattering: the gluons 
completely ``fill'' the target nucleus
\cite{Drescher2,Drescher:2004sd,Drescher:2005ak}. 
Not only the very numerous partons at small $x$ but also the much faster 
valence quarks will participate in the interaction. Indeed, in a
non-peripheral collision (complete saturation) the chance probability of 
valence quarks to scatter off these gluons approaches unity. As a
consequence this will lead to the disintegration of the leading valence
di-quark: no leading baryon is produced and the elasticity of the
collision drops by almost a factor 2.

Indeed there are some indications of ``anomalous'' baryon stopping in 
heavy ion collisions, however, at much lower energy
(for example, \cite{Mitchell:1993cm,Videbaek:2001mi}) 
and different theoretical interpretations are put forward
(for example, string junction interpretation: 
\cite{Capella:1996th,Capella:2002sx} and saturation
\cite{Itakura:2003jp}). 
Measurements of hadron production
in the very forward direction at RHIC \cite{Lange}, Tevatron
\cite{Rapidis}, and LHC \cite{Orava}
will be needed to study the leading baryon distributions systematically
and clarify the situation.

The implementation of different scenarios of parton density saturation
in the SIBYLL 2.1 code allows a first estimate of the expected effects
and their dependence on model 
parameters \cite{Drescher:2004sd,Drescher:2005ak}. 
In a conservative scenario the mean depth of shower maximum, 
$\langle X_{\rm max}\rangle$, is reduced
by about 20\,g/cm$^2$ at $10^{19}$\,eV, corresponding to
the difference between SIBYLL and QGSJET 01 predictions. Depending on the
assumptions, much larger reductions of $\langle X_{\rm max} \rangle$ are
possible.

%%%%%%%%%%%%%%%%%%%%%%%%%%%%%%%%%%%%%%%%%%%%%%%%%%%%%

\subsection{Hadronic interactions at intermediate energy}

Hadronic interactions in air showers at intermediate energy are often
simulated with models like 
GHEISHA~\cite{Fesefeldt85a} (used in CORSIKA \cite{Heck98a}) 
or the Hillas splitting
algorithm \cite{Hillas81a} (used in MOCCA \cite{Hillas95a} and AIRES
\cite{Sciutto:1999jh}).  Both models are very fast but rather crude 
parametrizations of low-energy data or interaction physics. 
Their application is certainly justified
if only the electron/photon component of a shower or calorimetric
quantities are studied as in this
case the details of low-energy interactions are of minor importance. 

The situation is, however, different for muons.
Due to successive hadronic interactions,
the number of pions and kaons increases in an air shower with decreasing
particle energy. Below some 100 (500) GeV pions (kaons) are more likely
to decay than to undergo further interactions.
Therefore, hadronic interactions in the energy range from several GeV to a few
hundred GeV are very 
important for understanding GeV muon production in EAS of all
energies \cite{Engel99c,Drescher:2002vp}. 
Recent studies have shown that the muon density at large 
lateral distance is indeed very sensitive to
the model used for low- and intermediate-energy interactions.
The differences between the predictions of the various
models are of the order of 10-20\% in the relevant lateral distance
range but can be as large as 50\% \cite{Drescher:2003gh,Heck:2003br}.

A detailed comparison of low- and intermediate-energy models to
available data \cite{Heck:2003br,Heck} show that
GHEISHA does not provide an adequate parametrization of the interaction
characteristics. The Hillas splitting algorithm seems to 
give a somewhat better
description but a thorough comparison to data is hampered by
the limitation to only p/$\pi$-air collisions.
The best models available are 
clearly FLUKA~\cite{Fasso01a} and UrQMD~\cite{Bleicher99a} but there are
still significant differences between the predictions of these two
models.

The energy range up to 400\,GeV is in reach of fixed-target accelerator
experiments. Not only that such experiments can easily measure
with light, air-like nuclei 
as targets they can also run with tagged pion and kaon beams
and measure particle production in the very forward direction.
Recognizing the importance of low-energy measurements
for atmospheric neutrino flux predictions \cite{Engel:1999zq} 
and neutrino factories, a
programme was begun to systematically measure pion and kaon production
in minimum bias collisions. Examples are the HARP experiment
\cite{Barr-Engel,Gomez-Cadenas:2004hy} where first data are available
now, the NA49 minimum bias p-C run \cite{Barr-Engel}, and the 
MIPP
experiment \cite{Rapidis,Raja:2005sh} which is currently taking data.

%%%%%%%%%%%%%%%%%%%%%%%%%%%%%%%%%%%%%%%%%%%%%%%%%%%%%

\subsection{Information from air shower measurements}

It is difficult to obtain information on hadronic multiparticle
production at ultra-high energy from EAS measurements. 
First of all the primary particle mass
is, in general, not known. Secondly, the large number of successive
hadronic interactions smears out any striking features of the primary
interaction. Therefore most analyses of air shower data in terms of
interaction physics are highly indirect and often serve only the exclusion of
extreme scenarios (for example, \cite{Drescher2}).

One of the most interesting measurements of this kind 
is the analysis of deeply penetrating
air showers to obtain the inelastic proton-air cross section. 
Traditionally an exponential function is fitted to the observed 
$X_{\rm max}$ distribution and the derived absorption length $\Lambda$
is converted to the interaction length via a so-called $k$-factor
(for a recent discussion, see \cite{Alvarez-Muniz:2004bx}).
The HiRes Collab.\ developed a new method that does partially avoid the
ambiguities of the definition of a $k$-factor \cite{Belov:2003ie}. 
Applying this method to the  stereo data set,
a preliminary p-air cross section of $456\pm 17{\rm (stat)} +
39 - 11 {\rm (sys)}$\,mb at $10^{18.5}$\,eV is derived \cite{Belov}. This
cross section is lower than the current model extrapolations, see
Fig.~\ref{fig:p-air-cs}. A number of possible biases still need to be
investigated. For example, a small fraction of photon primaries would be
enough to spoil the cross section measurement as photon-initiated
showers have a much larger depth of maximum. Furthermore the
self-consistency of the method should be checked by applying it to a 
model that is modified to match the actually measured cross section.

Experiments that measure many observables of air showers simultaneously
can check the consistency of EAS simulation. For example, the 
KASCADE installation allows the measurement of shower size
(electrons), muon
densities with thresholds of 240, 490, and 2400\,MeV, and hadron
multiplicities and energies above 70\,GeV in the shower 
core~\cite{Haungs1,Zabierowski}. 
The correlation between the different observables
provides constraints on interaction models even if the full
range of possible primary particles is considered \cite{Milke1}.
The latest versions of the had. interaction models available in CORSIKA 
satisfy these constraints but some earlier versions are clearly at
variance with the data. 

Emulsion chamber experiments at high altitude have the advantage that
they have a low shower energy threshold, reaching into 
the region of direct primary flux measurements. Therefore they are 
well suited for testing shower simulation models. For example, by
comparing the predicted and measured optical density distribution
(i.e. the energy distribution of particles in the TeV range) in
emulsions of the Pamir experiment, it was found
that some old versions of hadronic interaction models could be excluded
\cite{Haungs2,Haungs:2003bx}.

Finally it should be noted that 
there are some indications of a systematic discrepancy
between the mass composition derived from $N_e-N_\mu$ based measurements
and that from data sensitive to the depth of maximum
\cite{Hoerandel1,Hoerandel:2003vu}. 
It is unclear to what
extent these discrepancies are related to shortcomings in the simulation
of hadronic interactions, but one can try
to bring different measurements into better agreement by modifying the
underlying shower simulation correspondingly. A
hadronic interaction model with small cross section and a somewhat
reduced inelasticity as compared to QGSJET 01 is favoured in this analysis
\cite{Hoerandel:2003vu}.

%%%%%%%%%%%%%%%%%%%%%%%%%%%%%%%%%%%%%%%%%%%%%%%%%%%%%

\subsection{Trends in EAS simulation}

%%%%%%%%%%%%%%%%%%%%%%%%%%%%%%%%%%%%%%%%%%%
\begin{figure}[htb!]
\centerline{
\includegraphics[width=7.5cm]{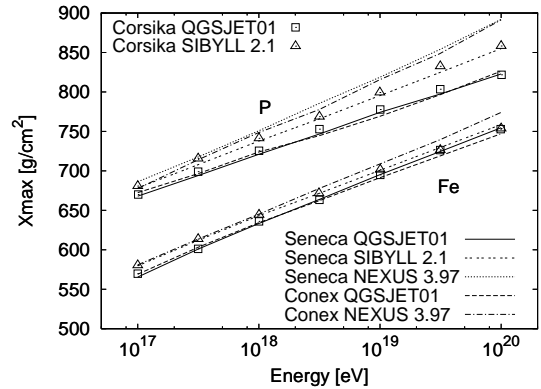}
}
\vspace*{-8mm}
\caption{
Mean depth of shower maximum for proton- and iron-induced showers
\cite{Drescher1}. 
The predictions of different hadronic interaction models as calculated
with different shower codes are compared.
\label{fig:xmax-models}
}
\end{figure}
%%%%%%%%%%%%%%%%%%%%%%%%%%%%%%%%%%%%%%%%%%%

There is a lack of statistics of Monte Carlo simulations in many
comparisons of data with theoretical predictions. For example,
the statistics of the KASCADE shower data exceed the 
simulated one available for each 
interaction model combination by about a factor 10. A similar disparity of
data to simulation statistics is also typical for modern 
large-scale detectors such as Auger. The problem is even amplified if
many observables are used to characterize each shower: the
set of simulated showers should by far exceed the number of observed
showers to keep the statistical reconstruction errors small.

At ultra-high energy, hybrid simulation schemes present a fast and efficient
alternative to conventional Monte Carlo simulation techniques.
In a hybrid simulation all interactions above a
certain energy threshold are simulated with the Monte Carlo technique.
Secondary particles that fall below this threshold are taken as sources
of subshowers that are treated numerically. The various hybrid schemes
available -- for a review see \cite{Drescher1} -- 
differ mainly in the method of calculating these sub-showers. 
For example, the sub-showers could be drawn from pre-calculated libraries
\cite{Alvarez-Muniz:2002ne}, calculated by solving cascade equations for
a Monte Carlo-generated source function \cite{Pierog}, or treated by
applying numerical solutions of cascade equations together with
analytical approximations or tables of shower evolution
\cite{Dedenko,Bossard:2000jh,Drescher:2002cr}.

The hybrid codes SENECA \cite{Drescher:2002cr} 
and CONEX \cite{Pierog} have several interaction models implemented and 
have reached a precision and sophistication that makes them suited
for analyzing experimental data \cite{Ortiz:2004gb}. SENECA allows a
full 3+1-dimensional simulation of EAS whereas CONEX is currently restricted to
calculating the projection along the shower axis. Both codes have been
extensively compared to CORSIKA. For example, a comparison of the mean
depth of maximum of different models is shown in
Fig.~\ref{fig:xmax-models}. The agreement between the different shower
simulation codes is excellent. It should also be noted that, only with
hybrid simulation codes, showers can be simulated in large numbers using
the time-consuming interaction model neXus 3.97 \cite{Pierog:2002gj}.

The simulation of inclined or even upward-going air showers has become
increasingly important. Large-aperture experiments like Auger and EUSO
have not only a large sensitivity to nearly horizontal showers but
also hope to find neutrino-induced upward-going showers
\cite{Kampert,Gorodetzky}. With the
exception of CONEX, the currently available shower simulation packages 
are not optimized for such calculations. Modifications and extensions 
will be needed to allow detailed and efficient simulation of showers at 
these particularly interesting geometries.

%%%%%%%%%%%%%%%%%%%%%%%%%%%%%%%%%%%%%%%%%%%%%%%%%%%%%%%%%%%%%%%%%%%%%%%

\section{Exotic phenomena and emulsion chamber data\label{sec:exotics}}

The most striking, unexpected phenomena observed in emulsion
chamber experiments are 
Centauro events with an exceptionally small number of photons,
events with particles or groups of particles being aligned
along a straight line,
halo events characterized by an unusually large area of darkness in
the X-ray film, and deeply penetrating cascades
\cite{Lattes:1980wk,Slavatinsky:2003zz}. Whether these phenomena are
related to fluctuations and the measurement technique of
emulsion chamber experiments or signs of new physics
is controversially debated for more than 30 years. 
In the following we will briefly discuss Centauro events and comment on
the status of searching for events with alignment 
(see \cite{Tamada} for a complete review).

There are
several aspects that complicate the interpretation of emulsion chamber
measurements.  First of
all, many of these phenomena are observed only in very high energy
events with estimated energy greater than $10^{16}$\,eV, of which  
only a small number of about 100 event has been collected
\cite{Slavatinsky:2003zz}. Secondly, due
to the threshold effect of the detectors ($E_\gamma > 1$\,TeV), mainly
proton initiated events are detected \cite{Haungs2}. 
As is well known, proton showers
are characterized by very large shower-to-shower fluctuations (see, for
example, \cite{Milke2}). Thirdly, the number of high energy
$\gamma$-rays and hadrons cannot be measured directly -- it is obtained by 
comparing the tracks at
various depths in the detector stacks.

\begin{figure}[htb!]
\centerline{
\includegraphics[width=7.5cm]{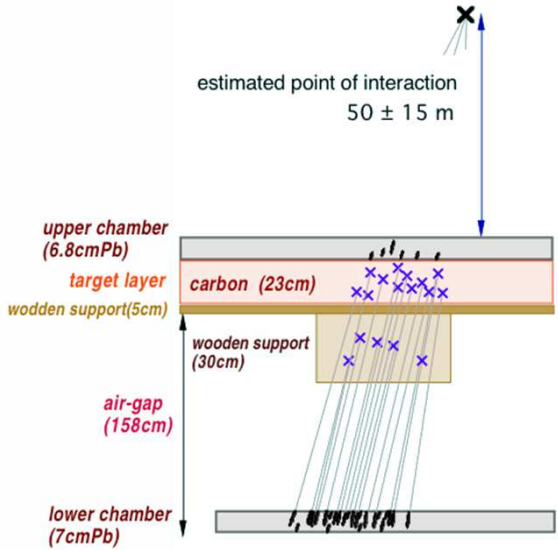}
}
\vspace*{-8mm}
\caption{
Illustration of the original interpretation of the Centauro-I event
\cite{Tamada}.
\label{fig:centauro}
}
\end{figure}

Probably the most famous exotic emulsion chamber event is Centauro-I, an
event with about 40 high-energy hadronic jets and only one low-energy 
e.m.{} cluster \cite{Lattes:1980wk}, see Fig.~\ref{fig:centauro}. 
Detected in 1972 in 
the Chacaltaya emulsion
chamber experiment \cite{Lattes:1973aa} 
it represents the most extreme Centauro event
ever observed. In total there are about 10 Centauro events observed
by the Chacaltaya and Pamir experiments (see \cite{Slavatinsky:2003zz}, a
detailed review of all events is given in
\cite{Gladysz-Dziadus:2001cq}). No
events of this kind were found in the experiments at Mount Fuji and
Mount Kanbala. Also all searches at accelerators were negative.

Models proposed for explaining Centauro events range from assuming a small
fraction of exotic primary
particles in the cosmic ray flux (for example, strangelets
\cite{Wlodarczyk,Rybczynski:2001bw} or
quark globs \cite{Bjorken:1979xv}) over exotic interaction scenarios, like
the creation of a disoriented chiral condensate \cite{Bjorken:1991xr}
or production of evaporating mini black holes by neutrino primaries
\cite{Tomaras,Mironov:2003jw}, to conventional features of 
diffraction dissociation \cite{Attallah:1993kc}.

%%%%%%%%%%%%%%%%%%%%%%%%%%%%%%%%%%%%%%%%%%%%
%%%%%%%%%%%%%%%%%%%%%%%%%%%%%%%%%%%%%%%%%%%%

Given these ongoing attempts of explaining Centauro events, the results
of the recent re-examination of the emulsion chamber plates are of
outstanding importance \cite{Tamada,Ohsawa:2004ta}. As it turned out,
the tracks in the two chambers that were previously thought to belong to the
Centauro-I event are actually due to two different events. The azimuth
angle of the tracks in the upper chamber does not match that 
in the lower one. As there is no counterpart found in the upper chamber,
this event is still very difficult to understand \cite{Ohsawa:2004ta}. 
The probability of 
particles produced in an interaction well above the installation passing
through a gap between the upper chambers seems to be very small.
A point-like interaction in the upper layers or the wooden support 
frame is also 
unlikely: the tracks would then correspond to very high $p_\perp$
particles and should point back to the interaction vertex. 
No such geometric convergence of the tracks is found.
This means that many particles of almost parallel trajectories have hit
the emulsion chamber without interacting in the upper lead stack,
again a very exotic scenario that lacks explanation.

The situation is similarly controversial regarding the experimental
information on events with co-planar particle emission. 

The Pamir Collab.\ find an excess of events with alignment
of the substructures for $E > 8 - 10 \times 10^{15}$\,eV
\cite{Borisov:2003th,Kopenkin:1994hu}. 
The highest energy event measured
with an emulsion experiment during a series of Concorde flights 
(average atm.\ depth $\sim 100$\,g/cm$^2$) shows also
impressive alignment \cite{Capdevielle:2001aa}.
At lower energy, no excess of coplanar particle production is found.
For example, measurements in the energy range
below $10^{14}$\,eV by the RUNJOB Collab.~\cite{Galkin:2001aa} 
and also a direct search with the CERN NA22 experiment at 
$2.5\times 10^{11}$\,eV \cite{Kopenkin:1994hu} 
provided distributions that agreed with the
expectations. Furthermore, a recent study of the KASCADE 
Collab.~\cite{Antoni:2005ce} showed that aligned
structures in hadronic shower cores at sea level are not related to
angular correlations in hadronic interactions as might be expected from
jet production \cite{Halzen:1989rg}. Indeed the fraction of events with
alignment is only determined by the lateral distribution of hadrons.
Measured in terms of the alignment parameter $\lambda_4$, the
event distributions of Pamir and KASCADE data look also surprisingly similar.

%%%%%%%%%%%%%%%%%%%%%%%%%%%%%%%%%%%%%%%%%%%%%%%%%%%%%%%%%%%%%%%%%%%%%%%

\section{Gamma-ray, neutrino, and muon flux
measurements\label{sec:secondaries}}

% In the following, gamma-rays, muons and neutrinos are considered as
% secondary particles as they are not accelerated directly but
% produced in the decay or interaction of hadrons. Of course, this applies
% only partially to gamma-rays: bremsstrahlung, synchrotron emission and
% inverse Compton scattering can also lead to high-energy gamma radiation.

Secondary particle fluxes such as hadronically produced gamma-rays and
neutrinos provide information on acceleration,
propagation and interaction of cosmic rays that are complementary to 
what can be directly deduced from the locally observed cosmic ray flux
\cite{DeRujula,Waxman,Berezinsky}.
In particular, gamma-rays and neutrinos propagate on straight
trajectories, allowing the identification of the source objects or
environments.

%%%%%%%%%%%%%%%%%%%%%%%%%%%%%%%%%%%%%%%%%%%%

\subsection{Gamma-rays}

With the begin of routine operation of the second
generation imaging atmospheric Cherenkov telescopes (IACT) 
CANGAROO~\cite{Kubo:2004ag}, 
HESS~\cite{Horns,Hinton:2004eu}, 
MAGIC~\cite{Fernandez,Bastieri:2005ry}, and 
VERITAS~\cite{Krennrich:2004ai},
many new TeV gamma-ray sources are
discovered and their spectra measured. It is impossible to summarize the
progress in this extremely active and diverse field of research 
and any comments made here will be soon outdated. 

At the time of writing this article all big four IACT installations are
completed and take data. Whereas CANGAROO and HESS have already
several telescopes online, MAGIC and VERITAS work 
with single, but bigger telescopes. 
Both the MAGIC and VERITAS
Collaborations are in the process of adding another 
telescope for stereoscopic
observation with greatly
improved background rejection. The HESS telescope system is
characterized by an unprecedented angular 
resolution of 0.06$^\circ$. The MAGIC telescope has a
light-weight design for very fast slew to observe transient sources.
It's low-energy threshold is planned to be about 20\,GeV as compared to
$\sim 70$\,GeV for HESS and VERITAS \cite{Fernandez}. 
The four telescopes together give almost uniform full-sky coverage.

Highlights of the early HESS data taking 
are certainly the measurement of the gamma-ray flux from
the Galactic Center \cite{Aharonian:2004wa}
and the first observation of a SNR as spatially
resolved TeV gamma-ray source (RXJ 1713-3946, a possible site of cosmic
ray acceleration) 
\cite{Aharonian:2004vr}. Both sources were previously detected
with the CANGAROO telescopes \cite{Tsuchiya:2004wv,Enomoto:2002xk} but
with much more limited resolution.
The potential of
the HESS telescopes is also underlined by the serendipituous discovery of an
unknown TeV gamma-ray source, now called TeV J1303-63, 
in the field of view of the binary pulsar
system PSR B1259-630 \cite{Horns}. 

In contrast to imaging Cherenkov telescopes, air shower arrays can be used to
continuously monitor the gamma-ray sky with very high duty cycle and
wide field of view. 
% They can also search for high energy gamma-rays
% coming from the direction of the sun. 
The two currently operated installations of this type
are Tibet AS$\gamma$~\cite{Ma1,Amenomori:2003zv} and
Milagro~\cite{Goodman,Sinnis:2003xv} being located at an altitude 
of 4300\,m and 2350\,m, respectively. Whereas the Milagro 
Collab.\ employ an active hadron/gamma-ray separation via two layers of PMTs in
an 8\,m deep water pond, the Tibet experiment searches for arrival direction
anisotropies due to gamma-rays on top of the isotropic cosmic ray
background with a dense scintillator array.

Both experiments have detected the Crab SNR and
the active galaxy Mrk 421 at the $5\sigma$ level
\cite{Atkins:2004yb,Amenomori:2005pn}. New results from Milagro are the 
detection of TeV gamma-rays from the entire inner Galactic plane 
region and the observation of two extended sources,
one coincident with EGRET source 3EG
J0520+2556 and another one in the Cygnus region of the Galactic plane
\cite{Atkins:2005wu,SazParkinson:2005td}. It is intriguing that the
latter source coincides within the Milagro resolution of about $2^\circ$ 
with the HEGRA source TeV J2032+4131 \cite{Aharonian:2005ex} and the region
from where AGASA reported an excess of $\sim 10^{18}$\,eV cosmic rays
\cite{Hayashida:1998qb}. At the moment an interpretation of these
observations in terms of a very high energy cosmic ray source (region) 
is too speculative -- more data will be needed.

%%%%%%%%%%%%%%%%%%%%%%%%%%%%%%%%%%%%%%%%%%%%

\subsection{High-energy neutrinos}

The interpretation of gamma-ray fluxes from potential cosmic ray sources
suffers from ambiguities due to the superposition of 
different gamma-ray production processes: $\pi^0$ decay, 
inverse Compton scattering, synchrotron radiation, and bremsstrahlung.
These uncertainties are expected to be much smaller for neutrinos as
they are
mainly produced in hadronic interactions via the decay of pions and kaons.
Furthermore neutrinos can travel over large distances virtually
unattenuated and are, therefore, ideal messenger particles, allowing a
multitude of astrophysical investigations \cite{Gaisser:1994yf}. On the other
hand, their small interaction cross section requires very large effective
detector volumes.

There are two neutrino telescopes taking data at the moment,
AMANDA-II \cite{Karle,Andres:1999hm} and 
Baikal NT-200 \cite{Tzamarias,Spiering:2004dt}. 
Although limited by detector size, the
sensitivity of both detectors will approach the cascade
bound in the next years\footnote{
The cascade bound, also called gamma-ray bound,
is based on the assumption that all
observed extragalactic gamma-rays were produced together with neutrinos in
hadronic cascades \cite{Berezinsky,Berezinsky:1975aa}.
}, i.e.\ touch the region where
one can hope for a discovery. The Waxman-Bahcall bound
\cite{Waxman:1998yy} (see also discussion
in \cite{Mannheim:1998wp}), often
considered as a reference flux that is guaranteed if protons are the
ultra-high energy cosmic rays, cannot be reached with these
installations.
About 3300 (370) neutrino candidates are found in the
AMANDA (Baikal) data taken until end of 2003. The number 
of neutrinos and their distribution are compatible with the
expectation from atmospheric neutrino production -- neither a diffuse flux
of extra-terrestial
neutrinos nor astrophysical point sources have yet been discovered.

The construction of the successor to AMANDA and much bigger neutrino 
telescope IceCube is on track \cite{Nygren}. 
IceCube will have a sensitivity that
reaches well into the region below the Waxman-Bahcall bound, promising
discoveries and many astrophysical and particle physics applications
\cite{Ahrens:2003ix}. The first IceCube string
was successfully deployed in February 2004 \cite{Nygren}.

The Mediterranean neutrino telescope collaborations
(ANTARES~\cite{Sokalski:2005sf},
NESTOR~\cite{Tsirigotis:2004bs}, NEMO~\cite{Migneco:2004yk})
\cite{Tzamarias} have performed prototype installations and test runs.
In 2003 the ANTARES Collab.\ operated a prototype sector line 
with PMTs and a mini
instrumentation line at the selected ANTARES site near Toulon (2500m
water depth). Valuable information on the performance of cables
and connectors under the harsh deep-sea conditions were gathered. It is
planned to build the complete ANTARES detector of 12 strings and 
in total 900 PMTs in 2005 -- 2007.
NEMO has selected a site close to Sicily with nearly perfect
conditions at 3500m water depth which is continuously monitored. 
The project is in the advanced R\&D
stage with the plan to build a prototype in 2005.
The NESTOR site provides a large plateau at the sea floor at about 4000 m
water depth. In 2003 one fully equipped prototype ``star'' of 32\,m
diameter of a 
NESTOR tower was successfully operated, allowing the measurement of the
atmospheric muon flux. It is planned to install 7 complete towers
by the end of 2006, providing a detector of about 0.15 km$^3$.

It is clear that a km$^3$-sized neutrino detector is needed
in the northern hemisphere to complement the field of view of IceCube.
Therefore the Mediterranean neutrino telescope collaborations
have recently joined their efforts to construct such a detector 
by initiating the design study KM3NeT \cite{KM3NeT}. 

To measure neutrino fluxes at even higher energy, radio emission of
neutrino-induced showers in dense materials can be employed \cite{Learned}.
Several experiments have recently performed measurements and derived
first limits on the neutrino flux at ultra-high energy 
(FORTE~\cite{Lehtinen:2003xv},
GLUE~\cite{Gorham:2003da},
ANITA~\cite{Miocinovic:2005jh}). 
For example, ANITA is designed to search for neutrinos with $E_\nu >
3\times 10^{18}$\,eV by monitoring radio signals from the antarctic ice
cap using a balloon-born system of antennas
\cite{Learned,Miocinovic:2005jh}.
A preparatory test flight with a prototype instrument (ANITA-lite) was
performed during the 03/04 austral season \cite{Learned}. 
Already on the basis of the prototype flight, giving about
7 days of data, a competitive 
limit on the ultra-high energy neutrino flux could be
derived \cite{Learned}.

%%%%%%%%%%%%%%%%%%%%%%%%%%%%%%%%%%%%%%%%%%%%

\subsection{High-energy muons}

Atmospheric muons, being a major background for neutrino
telescopes, carry valuable information as
messengers of hadronic interactions in the atmosphere. Muons are also
directly linked to atmospheric neutrino production and can be used to
test predictions on neutrino fluxes as needed for oscillation 
parameter analyses \cite{Gaisser:2002jj,Brancus}.
Particularly interesting is the comparison of muon fluxes to simulations
performed with the same codes as used for air shower analyses
\cite{LeCoultre,Ridky,Brancus,Ma2}. Of course, muon flux predictions
depend on both the used hadronic interaction model and the assumed
primary cosmic ray flux. 

%%%%%%%%%%%%%%%%%%%%%%%%%%%%%%%%
\begin{figure}[htb!]
\centerline{
\includegraphics[width=6.5cm]{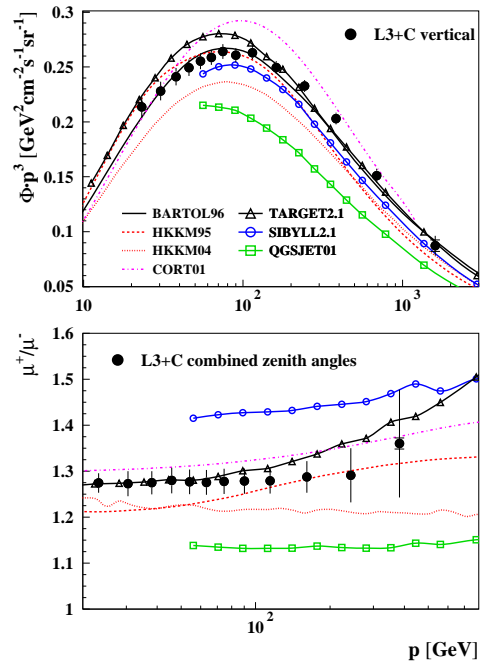}
}
\vspace*{-8mm}
\caption{Vertical atmospheric muon flux as measured by L3+C. The upper
panel shows the flux of all muons and the lower panel the charge ratio.
The data are compared with different theoretical predictions
\cite{LeCoultre,Achard:2004ws}. 
\label{fig:L3C-muons}
}
\end{figure}
%%%%%%%%%%%%%%%%%%%%%%%%%%%%%%%%

Very precise high-energy muon measurements can be carried out by 
particle physics detectors at colliders. For example, 
Fig.~\ref{fig:L3C-muons} shows the
inclusive muon flux and charge ratio measured by the L3 
Collab.~\cite{Achard:2004ws}. The experimental results for vertical muons are
compared to different theoretical predictions. In this case, in contrast to EAS
simulations, QGSJET01 predicts a smaller muon flux than 
SIBYLL 2.1 (cmp.\ Fig.~\ref{fig:ne-nmu-plot}). Hadronic interaction models
tuned for muon and neutrino flux calculations give a better description
of the data \cite{Ma2}. As known from simulations at lower energy
\cite{Brancus,Wentz:2003bp}, the muon charge ratio is found to be
very sensitive to the production of fast secondary hadrons. None of the
models implemented in CORSIKA gives a good overall description of the
data. The deficit of muons found relative to CORSIKA simulations
is also seen by two other LEP experiments. The DELPHI and Aleph groups find 
that the number of high energy muon bundles cannot be described by the
hadronic interaction models available in CORSIKA, assuming even a
completely iron-dominated primary composition
\cite{Ridky:2005mx,Avati:2000mn}.

At higher energy, the AMANDA and NESTOR collaborations have also 
measured the atmospheric muon flux in the
TeV energy region \cite{Karle,Tzamarias}. 
These measurements are integral flux determinations
because of the very limited energy resolution of  neutrino telescopes.
The AMANDA Collab.\ have also compared their measurement to simulations
with CORSIKA and find good agreement within the experimental
uncertainties \cite{Karle}.

%%%%%%%%%%%%%%%%%%%%%%%%%%%%%%%%%%%%%%%%%%%%%%%%%%%%%%%%%%%%%%%%%%%%%%%

\section{Conclusions and outlook\label{sec:outlook}}

Flux and composition of ultra-high energy cosmic rays are still very
uncertain because of the low statistics of showers observed so far and
the model-dependence of the shower data interpretation.  Significant
progress in this field is expected by new large-aperture installations
-- the Pierre Auger Observatory and the Telescope Array
\cite{TA}.  To study the flux at even higher energies $\sim10^{21}$\,eV
with sufficient statistics, new techniques will be required.  Observing
the atmosphere from outer space is one possible solution to increase the
aperture further (for example, EUSO \cite{Gorodetzky}). Another
possibility could be the use of radio antenna arrays similar to that of
particle detectors \cite{Falcke:2004aw}.

% With roughly half of the southern detector already completed, Auger has
% offered a fist glance at the correlation between longitudinal shower
% profile and particle densities at ground. No exceptional deviation is
% found.

The situation is similar in the knee energy region. Although the
all-particle flux is known rather well there are large uncertainties in
the composition. Nevertheless a clear trend from a mixed composition
at low energy to a predominantly heavy one above the knee is seen in all recent
measurements. The experimental errors are completely dominated by the
systematic uncertainties due to our limited understanding of hadronic
interactions at high energy and in forward direction.

The dependence on air shower simulations simulations can
be reduced by combining different detection techniques to measure
qualitatively different shower observables at the same time. 
Experiments of this
type are the Pierre Auger Observatory, TA, and IceCube/IceTop.

Whereas there are several detectors measuring the cosmic ray flux in the
knee region and at ultra-high energy, there is a lack of data in the
energy range $10^{17} - 10^{19}$\,eV. It is clear that the latter range
is of great interest as it expected to cover the transition from Galactic
to extragalactic cosmic rays. KASCADE-Grande and
IceTop will measure showers only up to $10^{17.5}$\,eV with good statistics.
Therefore it is worthwhile to upgrade large-aperture instruments such as
Auger or TA to extend their energy range down to $10^{17}$\,eV.

The field of emulsion chamber measurements is still full of mysteries.
After more than 30 years the interpretation of one of the most famous
emulsion chamber events, Centauro-I, has changed completely, now being
even more difficult to explain. Not a single non-emulsion experiment could
confirm any of the claimed exotic event features. Substantial progress
in this field can only be expected by new measurements combining
large-aperture emulsion stacks and modern particle detectors.

Many of the questions related to cosmic rays and astroparticle physics 
can only be solved by
measuring and understanding secondary particle fluxes. There has 
already been enormous progress in the field of gamma-ray 
and neutrino measurements and much more can be expected for the next
years. The second generation imaging air Cherenkov telescopes will provide
high resolution images of TeV gamma-ray sources and water/ice detectors 
of km$^3$ size will probe the neutrino flux in the same energy range.
Neutrino fluxes at ultra-high energy will be searched for by
large-aperture air shower installations and dedicated
radio signal experiments. The field of high-energy neutrino astronomy is
still in an infantile stage but with a bright future ahead.

For all these measurements and related data analyses, detailed
simulations are very important. Shower simulation tools in general and
hadronic interaction models in particular should be improved
continuously and tested by comparing them with a large variety of data.
During the last decade great progress was achieved by the introduction
of multi-purpose code packages such as CORSIKA and AIRES 
that are professionally maintained. 
However, it should not be overlooked that the quality of air
shower and inclusive flux simulations depends crucially on particle
production data measured in fixed-target and collider experiments. At
the moment the lack of suitable accelerator data is the dominant source of
systematic uncertainties in cosmic ray measurements. As we don't have a
calculable theory of hadronic multiparticle production, there is no
change of this dependence on accelerator data to be expected in near
future.

%%%%%%%%%%%%%%%%%%%%%%%%%%%%%%%%

\subsection*{Acknowledgements}

It is the author's pleasure to thank the organizers for inviting him to
participate in this
very interesting and fruitful symposium. He gratefully acknowledges
clarifying and illuminating discussions with many participants of this 
meeting and his colleagues from the KASCADE-Grande and Pierre Auger
collaborations. In particular he benefited from discussions with 
K.~Belov, V.~Berezinsky, A.~Haungs,  D.~Heck,
S.~Ostapchenko, T.~Pierog, L.~Resvanis, H.~Ulrich, M.~Unger, A.~Watson, 
and S.~Westerhoff. The author also would like to thank D.~Heck
for providing him Figs.~\ref{fig:ne-nmu-plot} and \ref{fig:p-air-cs}.

%%%%%%%%%%%%%%%%%%%%%%%%%%%%%%%%%%%%%%%%%%%%
% \bibliographystyle{unsrt-mod}
% \bibliographystyle{unsrt-mod-notitle}
% \bibliography{local,%

\begin{thebibliography}{100}

\small

\bibitem{Anchordoqui:2002hs}
L.~Anchordoqui, T.~Paul, S.~Reucroft, and J.~Swain,
Int. J. Mod. Phys. A18 (2003) 2229--2366
 and hep-ph/0206072.

\bibitem{Haungs:2003jv}
A.~Haungs, H.~Rebel, and M.~Roth,
Rept. Prog. Phys. 66 (2003) 1145--1206.

\bibitem{Watson:2003ba}
A.~A. Watson,
astro-ph/0312475.

\bibitem{Cronin:2004ye}
J.~W. Cronin,
astro-ph/0402487.

\bibitem{Engel:2004ui}
R.~Engel and H.~Klages,
Comptes Rendus Physique 5 (2004) 505--518.

\bibitem{Takeda:2003aa}
M.~Takeda {\it et~al.}  (AGASA Collab.),
Prepared for 28th International Cosmic Ray Conference (ICRC 2003), Tsukuba,
  Japan, 31 Jul - 7 Aug 2003, p. 381.

\bibitem{Shinozaki}
K.~Shinozaki  (AGASA Collab.),
these proceedings.

\bibitem{Nagano:1984db}
M.~Nagano {\it et~al.},
J. Phys. G10 (1984) 1295.

\bibitem{Nagano:1992jz}
M.~Nagano {\it et~al.},
J. Phys. G18 (1992) 423--442.

\bibitem{Westerhoff}
S.~Westerhoff  (HiRes Collab.),
these proceedings.

\bibitem{Abbasi:2002ta}
R.~U. Abbasi {\it et~al.}  (HiRes Collab.),
Phys. Rev. Lett. 92 (2004) 151101
 and astro-ph/0208243.

\bibitem{Abu-Zayyad:2002sf}
T.~Abu-Zayyad {\it et~al.}  (HiRes Collab.),
astro-ph/0208301.

\bibitem{Abu-Zayyad:2000ay}
T.~Abu-Zayyad {\it et~al.}  (HiRes-MIA Collab.),
Astrophys. J. 557 (2001) 686--699
 and astro-ph/0010652.

\bibitem{Haungs1}
A.~Haungs  (KASCADE Collab.),
these proceedings,
astro-ph/0412610.

\bibitem{Ulrich:2004bn}
H.~Ulrich {\it et~al.},
Eur. Phys. J. C33 (2004) s944--s946.

\bibitem{Fomin:2003tp}
Y.~A. Fomin {\it et~al.},
Prepared for 28th International Cosmic Ray Conference (ICRC 2003), Tsukuba,
  Japan, 31 Jul - 7 Aug 2003, p. 119-122.

\bibitem{Shibata}
T.~Shibata  (RUNJOB Collab.),
these proceedings.

\bibitem{Furukawa:2003dm}
M.~Furukawa {\it et~al.}  (RUNJOB Collab.),
Prepared for 28th International Cosmic Ray Conference (ICRC 2003), Tsukuba,
  Japan, 31 Jul - 7 Aug 2003, p. 1885-1888.

\bibitem{Ahn:2003cz}
H.~S. Ahn {\it et~al.}  (ATIC-1 Collab.),
Prepared for 28th International Cosmic Ray Conference (ICRC 2003), Tsukuba,
  Japan, 31 Jul - 7 Aug 2003 p. 1833-1836.

\bibitem{Navarra}
G.~Navarra  (EAS-TOP Collab.),
these proceedings.

\bibitem{Aglietta:1998te}
M.~Aglietta {\it et~al.}  (EAS-TOP Collab.),
Astropart. Phys. 10 (1999) 1--9.

\bibitem{Ma1}
Y.~Q. Ma  (Tibet AS$\gamma$ Collab.),
these proceedings.

\bibitem{Amenomori:2003tu}
M.~Amenomori {\it et~al.}  (Tibet AS$\gamma$ Collab.),
Prepared for 28th International Cosmic Ray Conference (ICRC 2003), Tsukuba,
  Japan, 31 Jul - 7 Aug 2003, p. 143-146.

\bibitem{Zatsepin}
G.~Zatsepin,
these proceedings.

\bibitem{Takeda:1998ps}
M.~Takeda {\it et~al.}  (AGASA Collab.),
Phys. Rev. Lett. 81 (1998) 1163--1166
 and astro-ph/9807193.

\bibitem{DeMarco:2003ig}
D.~De~Marco, P.~Blasi, and A.~V. Olinto,
Astropart. Phys. 20 (2003) 53--65
 and astro-ph/0301497.

\bibitem{Takeda:2002at}
M.~Takeda {\it et~al.}  (AGASA Collab.),
Astropart. Phys. 19 (2003) 447--462
 and astro-ph/0209422.

\bibitem{Bergman:2003aa}
D.~R. Bergman {\it et~al.}  (HiRes Collab.),
Prepared for 28th International Cosmic Ray Conference (ICRC 2003), Tsukuba,
  Japan, 31 Jul - 7 Aug 2003, p.~397.

\bibitem{Knurenko2}
S.~P. Knurenko  (Yakutsk Collab.),
these proceedings,
astro-ph/0411484.

\bibitem{Kalmykov92e}
N.~N. Kalmykov and S.~S. Ostapchenko,
Phys. At. Nucl. 56 (1993) (3) 346.

\bibitem{Kalmykov97a}
N.~N. Kalmykov, S.~Ostapchenko, and A.~I. Pavlov,
Nucl. Phys. B (Proc. Suppl.) 52B (1997) 17.

\bibitem{Abbasi:2004nz}
R.~U. Abbasi {\it et~al.}  (HiRes Collab.),
Astrophys. J. 622 (2005) 910--926
 and astro-ph/0407622.

\bibitem{Shinozaki:2004nh}
K.~Shinozaki and M.~Teshima  (AGASA Collab.),
Nucl. Phys. Proc. Suppl. 136 (2004) 18--27.

\bibitem{Knurenko1}
S.~P. Knurenko  (Yakutsk Collab.),
these proceedings,
astro-ph/0411483.

\bibitem{Bergman:2003wx}
D.~R. Bergman  (HiRes Collab.),
Mod. Phys. Lett. A18 (2003) 1235--1245
 and hep-ex/0307059.

\bibitem{Berezinsky:2005cq}
V.~Berezinsky, A.~Z. Gazizov, and S.~I. Grigorieva,
astro-ph/0502550.

\bibitem{Drescher2}
H.~J. Drescher,
these proceedings,
astro-ph/0411143.

\bibitem{Ostapchenko2}
S.~Ostapchenko,
these proceedings,
astro-ph/0412591.

\bibitem{Pierog}
T.~Pierog,
these proceedings,
astro-ph/0411260.

\bibitem{Fletcher94}
R.~S. Fletcher, T.~K. Gaisser, P.~Lipari, and T.~Stanev,
Phys. Rev. D50 (1994) 5710.

\bibitem{Engel99a}
R.~Engel, T.~K. Gaisser, P.~Lipari, and T.~Stanev,
in Proceedings of the 26th International Cosmic Ray Conference (Salt Lake City)
  vol.~1, p.~415,
1999.

\bibitem{Drescher:2000ha}
H.~J. Drescher, M.~Hladik, S.~Ostapchenko, T.~Pierog, and K.~Werner,
Phys. Rept. 350 (2001) 93--289
 and hep-ph/0007198.

\bibitem{Watson}
A.~A. Watson,
these proceedings,
astro-ph/0410514.

\bibitem{Ave:2003ab}
M.~Ave, J.~Knapp, M.~Marchesini, M.~Roth, and A.~A. Watson,
Prepared for 28th International Cosmic Ray Conference (ICRC 2003), Tsukuba,
  Japan, 31 Jul - 7 Aug 2003, p.~349.

\bibitem{Dova:2004nq}
M.~T. Dova, M.~E. Mancenido, A.~G. Mariazzi, T.~P. McCauley, and A.~A. Watson,
Astropart. Phys. 21 (2004) 597--607.

\bibitem{Homola}
P.~Homola,
these proceedings,
astro-ph/0411060.

\bibitem{Risse:2005jr}
M.~Risse {\it et~al.},
astro-ph/0502418.

\bibitem{Teshima:2003ab}
M.~Teshima {\it et~al.}  (AGASA Collab.),
Prepared for 28th International Cosmic Ray Conference (ICRC 2003), Tsukuba,
  Japan, 31 Jul - 7 Aug 2003, p.~437.

\bibitem{Finley:2003ur}
C.~B. Finley and S.~Westerhoff,
Astropart. Phys. 21 (2004) 359--367
 and astro-ph/0309159.

\bibitem{Stanev:1995my}
T.~Stanev, P.~L. Biermann, J.~Lloyd-Evans, J.~P. Rachen, and A.~Watson,
Phys. Rev. Lett. 75 (1995) 3056--3059
 and astro-ph/9505093.

\bibitem{Tinyakov:2001nr}
P.~G. Tinyakov and I.~I. Tkachev,
JETP Lett. 74 (2001) 445--448
 and astro-ph/0102476.

\bibitem{Gorbunov:2002hk}
D.~S. Gorbunov, P.~G. Tinyakov, I.~I. Tkachev, and S.~V. Troitsky,
Astrophys. J. 577 (2002) L93
 and astro-ph/0204360.

\bibitem{Tinyakov:2001ir}
P.~G. Tinyakov and I.~I. Tkachev,
Astropart. Phys. 18 (2002) 165--172
 and astro-ph/0111305.

\bibitem{Gorbunov:2004bs}
D.~S. Gorbunov, P.~G. Tinyakov, I.~I. Tkachev, and S.~V. Troitsky,
JETP Lett. 80 (2004) 145--148
 and astro-ph/0406654.

\bibitem{Kampert}
K.~H. Kampert  (Auger Collab.),
these proceedings,
astro-ph/0501074.

\bibitem{Auger}
http://www.auger.org.

\bibitem{Fukushima:2003ig}
M.~Fukushima,
Prog. Theor. Phys. Suppl. 151 (2003) 206--210.

\bibitem{TA}
http://www-ta.icrr.u-tokyo.ac.jp/TA\_Proposal/.

\bibitem{Abraham:2004dt}
J.~Abraham {\it et~al.}  (Auger Collab.),
Nucl. Instrum. Meth. A523 (2004) 50--95.

\bibitem{Hoerandel:2004gv}
J.~R. Hoerandel,
Astropart. Phys. 21 (2004) 241--265
 and astro-ph/0402356.

\bibitem{Swordy:2002df}
S.~P. Swordy {\it et~al.},
Astropart. Phys. 18 (2002) 129--150
 and astro-ph/0202159.

\bibitem{Karle}
A.~Karle  (AMANDA Collab.),
these proceedings.

\bibitem{Petrukhin}
A.~A. Petrukhin,
these proceedings.

\bibitem{Antoni:2002au}
T.~Antoni {\it et~al.}  (KASCADE Collab.),
Astropart. Phys. 16 (2002) 373--386.

\bibitem{Antoni:2003gd}
T.~Antoni {\it et~al.}  (KASCADE Collab.),
Nucl. Instrum. Meth. A513 (2003) 490--510.

\bibitem{Aglietta:2003hq}
M.~Aglietta {\it et~al.}  (MACRO Collab.),
Astropart. Phys. 20 (2004) 641--652
 and astro-ph/0305325.

\bibitem{Ahrens:2004nn}
J.~Ahrens {\it et~al.}  (AMANDA and SPASE Collab.),
Astropart. Phys. 21 (2004) 565--581.

\bibitem{Amenomori:2003tk}
M.~Amenomori {\it et~al.}  (Tibet AS$\gamma$ Collab.),
Prepared for 28th International Cosmic Ray Conference (ICRC 2003), Tsukuba,
  Japan, 31 Jul - 7 Aug 2003, p. 107-110.

\bibitem{Heck98a}
D.~Heck, J.~Knapp, J.~Capdevielle, G.~Schatz, and T.~Thouw,
in Wissenschaftliche Berichte FZKA 6019, Forschungszentrum Karlsruhe,
1998.

\bibitem{Asakimori:1998aa}
K.~Asakimori {\it et~al.}  (JACEE Collab.),
Astrophys. J. 502 (1998) 278--283.

\bibitem{Ahn:2003de}
H.~S. Ahn {\it et~al.}  (ATIC-2 Collab.),
Prepared for 28th International Cosmic Ray Conference (ICRC 2003), Tsukuba,
  Japan, 31 Jul - 7 Aug 2003, 1853-1856.

\bibitem{Candia:2003dk}
J.~Candia, S.~Mollerach, and E.~Roulet,
JCAP 0305 (2003) 003
 and astro-ph/0302082.

\bibitem{Antoni:2003jm}
T.~Antoni {\it et~al.}  (KASCADE Collab.),
Astrophys. J. 604 (2004) 687--692
 and astro-ph/0312375.

\bibitem{Antoni:2004sc}
T.~Antoni {\it et~al.}  (KASCADE Collab.),
Astrophys. J. 608 (2004) 865--871
 and astro-ph/0402656.

\bibitem{Amenomori:2004bf}
M.~Amenomori {\it et~al.}  (Tibet AS$\gamma$ Collab.),
Phys. Rev. Lett. 93 (2004) 061101
 and astro-ph/0408187.

\bibitem{Aglietta:1996sz}
M.~Aglietta {\it et~al.}  (EAS-TOP Collab.),
Astrophys. J. 470 (1996) 501--505.

\bibitem{Antoni:2001pw}
T.~Antoni {\it et~al.}  (KASCADE Collab.),
Astropart. Phys. 16 (2002) 245--263
 and astro-ph/0102443.

\bibitem{Knapp96a}
J.~Knapp, D.~Heck, and G.~Schatz,
in Wissenschaftliche Berichte FZKA 5828, Forschungszentrum Karlsruhe,
1996.

\bibitem{Knapp:2002vs}
J.~Knapp, D.~Heck, S.~J. Sciutto, M.~T. Dova, and M.~Risse,
Astropart. Phys. 19 (2003) 77--99
 and astro-ph/0206414.

\bibitem{Engel99c}
R.~Engel, T.~K. Gaisser, and T.~Stanev,
Proc. of XXIX Int. Symposium on Multiparticle Dynamics, Brown Univerity, USA,
  Aug. 8-13, p. 457,
2000.

\bibitem{Drescher:2002vp}
H.-J. Drescher and G.~R. Farrar,
Astropart. Phys. 19 (2003) 235--244
 and hep-ph/0206112.

\bibitem{Heck:2003br}
D.~Heck {\it et~al.},
Prepared for 28th International Cosmic Ray Conference (ICRC 2003), Tsukuba,
  Japan, 31 Jul - 7 Aug 2003, p.~279.

\bibitem{Haenssgen86a}
K.~H\"an{\ss}gen and J.~Ranft,
Comp. Phys. Commun. 39 (1986) 37.

\bibitem{Muecke00a}
A.~M\"ucke, R.~Engel, R.~J. Protheroe, J.~P. Rachen, and T.~Stanev,
Comp. Phys. Commun. 124 (2000) 290.

\bibitem{Fasso01a}
A.~Fasso, A.~Ferrari, J.~Ranft, and R.~P. Sala,
in Proc. of Int. Conf. on Advanced Monte Carlo for Radiation Physics, Particle
  Transport Simulation and Applications (MC 2000), Lisbon, Portugal, 23-26 Oct
  2000, A. Kling, F. Barao, M. Nakagawa, L. Tavora, P. Vaz eds.,
  Springer-Verlag Berlin, p. 955,
2001.

\bibitem{Ranft95a}
J.~Ranft,
Phys. Rev. D51 (1995) 64.

\bibitem{Heck-Ostapchenko-private}
D.~Heck and S.~Ostapchenko,
private communication.

\bibitem{Engel03a}
R.~Engel,
Nucl. Phys. B (Proc. Suppl.) 122 (2003) 40.

\bibitem{Alvarez-Muniz:2002ne}
J.~Alvarez-Muniz, R.~Engel, T.~K. Gaisser, J.~A. Ortiz, and T.~Stanev,
Phys. Rev. D66 (2002) 033011
 and astro-ph/0205302.

\bibitem{Ostapchenko:2003sj}
S.~S. Ostapchenko,
J. Phys. G29 (2003) 831--842.

\bibitem{Stanev}
T.~Stanev,
these proceedings.

\bibitem{Gribov83}
V.~N. Gribov, M.~L. Levin, and M.~G. Ryskin,
Phys. Rep. 100 (1983) 1.

\bibitem{Mueller:2005me}
A.~H. Mueller,
hep-ph/0501012.

\bibitem{Erice04}
talks at Workshop ``QCD at Cosmic Energies'', Aug 29 - Sep 5, 2004, Erice,
  Italy, http://www.lpthe.jussieu.fr/Erice/, the next workshop will be held
  2005 in Greece, see http://www.lpthe.jussieu.fr/Greece/.

\bibitem{Iancu:2003xm}
E.~Iancu and R.~Venugopalan,
hep-ph/0303204.

\bibitem{Golec-Biernat:1998js}
K.~Golec-Biernat and M.~Wusthoff,
Phys. Rev. D59 (1999) 014017
 and hep-ph/9807513.

\bibitem{Forshaw:2004rn}
J.~R. Forshaw, R.~Sandapen, and G.~Shaw,
Int. J. Mod. Phys. A19 (2004) 5425--5432
 and hep-ph/0407261.

\bibitem{Abramowicz99a}
H.~Abramowicz and A.~Caldwell,
Rev. Mod. Phys. 71 (1999) 1275.

\bibitem{Lange}
S.~Lange,
these proceedings.

\bibitem{Levin:2004ak}
E.~Levin,
hep-ph/0408039.

\bibitem{Steinberg:2004ii}
P.~A. Steinberg,
Acta Phys. Polon. B35 (2004) 235--239.

\bibitem{Kaidalov:1983vn}
A.~B. Kaidalov and K.~A. Ter-Martirosian,
Sov. J. Nucl. Phys. 39 (1984) 979.

\bibitem{Ostapchenko1}
S.~Ostapchenko,
these proceedings,
astro-ph/0412332.

\bibitem{Kaidalov86b-e}
A.~B. Kaidalov, L.~A. Ponomarev, and K.~A. Ter-Martirosyan,
Sov. J. Nucl. Phys. 44 (1986) 468.

\bibitem{Luna:2004eg}
R.~Luna, A.~Zepeda, C.~A. Garcia~Canal, and S.~J. Sciutto,
Phys. Rev. D70 (2004) 114034
 and hep-ph/0408303.

\bibitem{Engel:2002id}
R.~Engel,
Nucl. Phys. Proc. Suppl. 122 (2003) 437--446
 and hep-ph/0212340.

\bibitem{Heck-private}
D.~Heck,
private communication.

\bibitem{Rapidis}
P.~Rapidis,
these proceedings.

\bibitem{Albrow99a}
M.~Albrow, L.~Nodulman, and P.~Giromini,
CDF/PUBLIC/4844, available from CDF web page,
1999.

\bibitem{Zha:2003bt}
M.~Zha, J.~Knapp, and S.~Ostapchenko,
Prepared for 28th International Cosmic Ray Conference (ICRC 2003), Tsukuba,
  Japan, 31 Jul - 7 Aug 2003, p. 515-518.

\bibitem{Orava}
R.~Orava,
these proceedings.

\bibitem{Avati:2003qj}
V.~Avati {\it et~al.}  (TOTEM Collab.),
Eur. Phys. J. C34 (2004) s255--s268.

\bibitem{Engel:1998hf}
R.~Engel,
Nucl. Phys. B (Proc. Suppl.) 75A (1999) 62
 and astro-ph/9811225.

\bibitem{Frankfurt:1997ij}
L.~Frankfurt, W.~Koepf, and M.~Strikman,
Phys. Lett. B405 (1997) 367--372
 and hep-ph/9702236.

\bibitem{Drescher:2004sd}
H.~J. Drescher, A.~Dumitru, and M.~Strikman,
hep-ph/0408073.

\bibitem{Drescher:2005ak}
H.~J. Drescher, A.~Dumitru, and M.~Strikman,
hep-ph/0501165.

\bibitem{Mitchell:1993cm}
J.~T. Mitchell  (NA35 Collab.),
Nucl. Phys. A566 (1994) 415c--418c.

\bibitem{Videbaek:2001mi}
F.~Videbaek,
Heavy Ion Phys. 15 (2002) 303--313
 and nucl-ex/0106017.

\bibitem{Capella:1996th}
A.~Capella and B.~Z. Kopeliovich,
Phys. Lett. B381 (1996) 325--330
 and hep-ph/9603279.

\bibitem{Capella:2002sx}
A.~Capella,
Phys. Lett. B542 (2002) 65--70.

\bibitem{Itakura:2003jp}
K.~Itakura, Y.~V. Kovchegov, L.~McLerran, and D.~Teaney,
Nucl. Phys. A730 (2004) 160--190
 and hep-ph/0305332.

\bibitem{Fesefeldt85a}
H.~Fesefeldt,
RWTH Aachen,
1985.

\bibitem{Hillas81a}
A.~M. Hillas,
in Proc. 17th Int. Cosmic Ray Conf. {\bf 8}, p.~193, Paris, France,
1981.

\bibitem{Hillas95a}
A.~M. Hillas,
in Proc. 24th Int. Cosmic Ray Conf. {\bf 1}, p.~270, Rome, Italy,
1995.

\bibitem{Sciutto:1999jh}
S.~J. Sciutto,
astro-ph/9911331.

\bibitem{Drescher:2003gh}
H.-J. Drescher, M.~Bleicher, S.~Soff, and H.~Stoecker,
Astropart. Phys. 21 (2004) 87--94
 and astro-ph/0307453.

\bibitem{Heck}
D.~Heck,
private communication,
astro-ph/0410735.

\bibitem{Bleicher99a}
M.~Bleicher {\it et~al.},
J. Phys. G: Nucl. Part. Phys. 25 (1999) 1859.

\bibitem{Engel:1999zq}
R.~Engel, T.~K. Gaisser, and T.~Stanev,
Phys. Lett. B472 (2000) 113--118
 and hep-ph/9911394.

\bibitem{Barr-Engel}
G.~Barr and R.~Engel,
these proceedings,
astro-ph/0404356.

\bibitem{Gomez-Cadenas:2004hy}
J.~J. Gomez-Cadenas  (HARP Collab.),
Nucl. Phys. Proc. Suppl. 143 (2005) 291--296
 and hep-ex/0410043.

\bibitem{Raja:2005sh}
R.~Raja,
hep-ex/0501005,
2005.

\bibitem{Alvarez-Muniz:2004bx}
J.~Alvarez-Muniz, R.~Engel, T.~K. Gaisser, J.~A. Ortiz, and T.~Stanev,
Phys. Rev. D69 (2004) 103003
 and astro-ph/0402092.

\bibitem{Belov:2003ie}
K.~Belov  (HiRes Collab.),
Prepared for 28th International Cosmic Ray Conference (ICRC 2003), Tsukuba,
  Japan, 31 Jul - 7 Aug 2003, p. 1567-1570.

\bibitem{Belov}
K.~Belov  (HiRes Collab.),
these proceedings.

\bibitem{Zabierowski}
J.~Zabierowski  (KASCADE Collab.),
these proceedings.

\bibitem{Milke1}
J.~Milke  (KASCADE Collab.),
these proceedings.

\bibitem{Haungs2}
A.~Haungs,
these proceedings.

\bibitem{Haungs:2003bx}
A.~Haungs and J.~Kempa,
Nuovo Cim. 26C (2003) 503--520.

\bibitem{Hoerandel1}
J.~R. Hoerandel,
these proceedings.

\bibitem{Hoerandel:2003vu}
J.~R. Hoerandel,
J. Phys. G29 (2003) 2439--2464
 and astro-ph/0309010.

\bibitem{Drescher1}
H.~J. Drescher,
these proceedings,
astro-ph/0411144.

\bibitem{Dedenko}
L.~G. Dedenko,
these proceedings.

\bibitem{Bossard:2000jh}
G.~Bossard {\it et~al.},
Phys. Rev. D63 (2001) 054030
 and hep-ph/0009119.

\bibitem{Drescher:2002cr}
H.-J. Drescher and G.~R. Farrar,
Phys. Rev. D67 (2003) 116001
 and astro-ph/0212018.

\bibitem{Ortiz:2004gb}
J.~A. Ortiz, G.~A. Medina~Tanco, and V.~de~Souza,
astro-ph/0411421.

\bibitem{Pierog:2002gj}
T.~Pierog, H.~J. Drescher, F.~Liu, S.~Ostaptchenko, and K.~Werner,
Nucl. Phys. A715 (2003) 895--898
 and hep-ph/0211202.

\bibitem{Gorodetzky}
P.~Gorodetzky  (EUSO Collab.),
these proceedings,
astro-ph/0502187.

\bibitem{Lattes:1980wk}
C.~M.~G. Lattes, Y.~Fujimoto, and S.~Hasegawa,
Phys. Rept. 65 (1980) 151.

\bibitem{Slavatinsky:2003zz}
S.~A. Slavatinsky,
Nucl. Phys. Proc. Suppl. 122 (2003) 3--11.

\bibitem{Tamada}
M.~Tamada,
these proceedings.

\bibitem{Milke2}
J.~Milke  (KASCADE Collab.),
these proceedings.

\bibitem{Lattes:1973aa}
C.~M.~G. Lattes {\it et~al.}  (Brazil-Japan Collab.),
Prepared for 13th International Cosmic Ray Conference (ICRC 1973), Denver, CO,
  17 - 30 Aug 1973, vol. 3, p. 2227; vol. 4, p. 2671.

\bibitem{Gladysz-Dziadus:2001cq}
E.~Gladysz-Dziadus,
Phys. Part. Nucl. 34 (2003) 285--347
 and hep-ph/0111163.

\bibitem{Wlodarczyk}
Z.~Wlodarczyk,
these proceedings,
hep-ph/0410064.

\bibitem{Rybczynski:2001bw}
M.~Rybczynski, Z.~Wlodarczyk, and G.~Wilk,
Acta Phys. Polon. B33 (2002) 277--296
 and hep-ph/0109225.

\bibitem{Bjorken:1979xv}
J.~D. Bjorken and L.~D. McLerran,
Phys. Rev. D20 (1979) 2353.

\bibitem{Bjorken:1991xr}
J.~D. Bjorken,
Int. J. Mod. Phys. A7 (1992) 4189--4258.

\bibitem{Tomaras}
T.~N. Tomaras,
these proceedings,
hep-ph/0411081.

\bibitem{Mironov:2003jw}
A.~Mironov, A.~Morozov, and T.~N. Tomaras,
hep-ph/0311318.

\bibitem{Attallah:1993kc}
R.~Attallah and J.~N. Capdevielle,
J. Phys. G19 (1993) 1381--1392.

\bibitem{Ohsawa:2004ta}
A.~Ohsawa, E.~H. Shibuya, and M.~Tamada,
Phys. Rev. D70 (2004) 074028.

\bibitem{Borisov:2003th}
A.~S. Borisov, V.~M. Maksimenko, R.~A. Mukhamedshin, V.~S. Puchkov, and S.~A.
  Slavatinsky,
Prepared for 28th International Cosmic Ray Conference (ICRC 2003), Tsukuba,
  Japan, 31 Jul - 7 Aug 2003, p. 85-88.

\bibitem{Kopenkin:1994hu}
V.~V. Kopenkin, A.~K. Managadze, I.~V. Rakobolskaya, and T.~M. Roganova,
Phys. Rev. D52 (1995) 2766--2774
 and hep-ph/9408247.

\bibitem{Capdevielle:2001aa}
J.~N. Capdevielle, R.~Attallah, and M.~C. Talai,
Prepared for 27th International Cosmic Ray Conference (ICRC 2001), Hamburg,
  Germany, 7-15 Aug 2001, p. 1410-1413.

\bibitem{Galkin:2001aa}
V.~I. Galkin {\it et~al.},
Prepared for 27th International Cosmic Ray Conference (ICRC 2001), Hamburg,
  Germany, 7-15 Aug 2001, p. 1407-1409.

\bibitem{Antoni:2005ce}
T.~Antoni {\it et~al.}  (KASCADE Collab.),
Phys. Rev. D71 (2005) 072002
 and hep-ph/0503218.

\bibitem{Halzen:1989rg}
F.~Halzen and D.~A. Morris,
Phys. Rev. D42 (1990) 1435--1439.

\bibitem{DeRujula}
A.~De~R\'ujula,
these proceedings,
astro-ph/0412094.

\bibitem{Waxman}
E.~Waxman,
these proceedings,
astro-ph/0412554.

\bibitem{Berezinsky}
V.~Berezinsky,
these proceedings.

\bibitem{Kubo:2004ag}
H.~Kubo {\it et~al.}  (CANGAROO Collab.),
New Astron. Rev. 48 (2004) 323--329.

\bibitem{Horns}
D.~Horns,
these proceedings.

\bibitem{Hinton:2004eu}
J.~A. Hinton  (HESS Collab.),
New Astron. Rev. 48 (2004) 331--337
 and astro-ph/0403052.

\bibitem{Fernandez}
E.~Fern\'andez,
these proceedings.

\bibitem{Bastieri:2005ry}
D.~Bastieri {\it et~al.}  (MAGIC Collab.),
astro-ph/0503534.

\bibitem{Krennrich:2004ai}
F.~Krennrich {\it et~al.},
New Astron. Rev. 48 (2004) 345--349.

\bibitem{Aharonian:2004wa}
F.~Aharonian {\it et~al.}  (HESS Collab.),
astro-ph/0408145.

\bibitem{Aharonian:2004vr}
F.~A. Aharonian {\it et~al.}  (HESS Collab.),
Nature. 432 (2004) 75--77
 and astro-ph/0411533.

\bibitem{Tsuchiya:2004wv}
K.~Tsuchiya {\it et~al.}  (CANGAROO-II Collab.),
Astrophys. J. 606 (2004) L115--L118
 and astro-ph/0403592.

\bibitem{Enomoto:2002xk}
R.~Enomoto {\it et~al.},
Nature 416 (2002) 823--826.

\bibitem{Amenomori:2003zv}
M.~Amenomori {\it et~al.}  (Tibet AS$\gamma$ Collab.),
Prepared for 28th International Cosmic Ray Conference (ICRC 2003), Tsukuba,
  Japan, 31 Jul - 7 Aug 2003, p. 3019-3022.

\bibitem{Goodman}
J.~Goodman  (Milagro Collab.),
these proceedings.

\bibitem{Sinnis:2003xv}
G.~Sinnis  (Milagro Collab.),
Prepared for 28th International Cosmic Ray Conference (ICRC 2003), Tsukuba,
  Japan, 31 Jul - 7 Aug 2003, p. 2583-2586.

\bibitem{Atkins:2004yb}
R.~Atkins {\it et~al.},
Astrophys. J. 608 (2004) 680--685
 and astro-ph/0403097.

\bibitem{Amenomori:2005pn}
M.~Amenomori {\it et~al.}  (Tibet AS$\gamma$ Collab.),
astro-ph/0502039.

\bibitem{Atkins:2005wu}
A.~Atkins {\it et~al.}  (Milagro Collab.),
astro-ph/0502303.

\bibitem{SazParkinson:2005td}
P.~M. Saz~Parkinson  (Milagro Collab.),
astro-ph/0503244.

\bibitem{Aharonian:2005ex}
F.~Aharonian {\it et~al.}  (HEGRA Collab.),
astro-ph/0501667.

\bibitem{Hayashida:1998qb}
N.~Hayashida {\it et~al.}  (AGASA Collab.),
Astropart. Phys. 10 (1999) 303--311
 and astro-ph/9807045.

\bibitem{Gaisser:1994yf}
T.~K. Gaisser, F.~Halzen, and T.~Stanev,
Phys. Rept. 258 (1995) 173--236
 and hep-ph/9410384.

\bibitem{Andres:1999hm}
E.~Andres {\it et~al.}  (AMANDA Collab.),
Astropart. Phys. 13 (2000) 1--20
 and astro-ph/9906203.

\bibitem{Tzamarias}
S.~Tzamarias,
these proceedings.

\bibitem{Spiering:2004dt}
C.~Spiering {\it et~al.}  (BAIKAL Collab.),
astro-ph/0404096.

\bibitem{Berezinsky:1975aa}
V.~S. Berezinsky and A.~Y. Smirnov,
Astrop. Sp. Sci. 32 (1975) 461.

\bibitem{Waxman:1998yy}
E.~Waxman and J.~N. Bahcall,
Phys. Rev. D59 (1999) 023002
 and hep-ph/9807282.

\bibitem{Mannheim:1998wp}
K.~Mannheim, R.~J. Protheroe, and J.~P. Rachen,
Phys. Rev. D63 (2001) 023003
 and astro-ph/9812398.

\bibitem{Nygren}
D.~Nygren  (IceCube Collab.),
these proceedings.

\bibitem{Ahrens:2003ix}
J.~Ahrens {\it et~al.}  (IceCube Collab.),
Astropart. Phys. 20 (2004) 507--532
 and astro-ph/0305196.

\bibitem{Sokalski:2005sf}
I.~Sokalski  (ANTARES Collab.),
hep-ex/0501003.

\bibitem{Tsirigotis:2004bs}
A.~G. Tsirigotis  (NESTOR Collab.),
Eur. Phys. J. C33 (2004) s956--s958.

\bibitem{Migneco:2004yk}
E.~Migneco {\it et~al.},
To appear in the proceedings of 21st International Conference on Neutrino
  Physics and Astrophysics (Neutrino 2004), Paris, France, 14-19 Jun 2004.

\bibitem{KM3NeT}
http://km3net.org.

\bibitem{Learned}
J.~Learned,
these proceedings.

\bibitem{Lehtinen:2003xv}
N.~G. Lehtinen, P.~W. Gorham, A.~R. Jacobson, and R.~A. Roussel-Dupre,
Phys. Rev. D69 (2004) 013008
 and astro-ph/0309656.

\bibitem{Gorham:2003da}
P.~W. Gorham {\it et~al.},
Phys. Rev. Lett. 93 (2004) 041101
 and astro-ph/0310232.

\bibitem{Miocinovic:2005jh}
P.~Miocinovic {\it et~al.},
astro-ph/0503304.

\bibitem{Gaisser:2002jj}
T.~K. Gaisser and M.~Honda,
Ann. Rev. Nucl. Part. Sci. 52 (2002) 153--199
 and hep-ph/0203272.

\bibitem{Brancus}
I.~Brancus,
these proceedings.

\bibitem{LeCoultre}
P.~Le~Coultre  (L3 Collab.),
these proceedings.

\bibitem{Ridky}
J.~Ridky,
these proceedings.

\bibitem{Ma2}
Y.~Q. Ma,
these proceedings.

\bibitem{Achard:2004ws}
P.~Achard {\it et~al.}  (L3 Collab.),
Phys. Lett. B598 (2004) 15--32
 and hep-ex/0408114.

\bibitem{Wentz:2003bp}
J.~Wentz {\it et~al.},
Phys. Rev. D67 (2003) 073020
 and hep-ph/0301199.

\bibitem{Ridky:2005mx}
J.~Ridky and P.~Travnicek  (DELPHI Collab.),
Nucl. Phys. Proc. Suppl. 138 (2005) 295--298.

\bibitem{Avati:2000mn}
V.~Avati {\it et~al.},
Astropart. Phys. 19 (2003) 513--523.

\bibitem{Falcke:2004aw}
H.~Falcke, P.~Gorham, and R.~J. Protheroe,
New Astron. Rev. 48 (2004) 1487--1510
 and astro-ph/0409229.

\end{thebibliography}
%               /home/engel/tex/hep-2,%
%               /home/engel/tex/astro-2,%
%               /home/engel/tex/p-nuc-data-1}
%%%%%%%%%%%%%%%%%%%%%%%%%%%%%%%%%%%%%%%%%%%%

\end{document}